\documentclass[fleqn,10pt]{wlscirep}
\usepackage[utf8]{inputenc}
\usepackage{units}
\usepackage{makecell}
\usepackage{float}
\usepackage[T1]{fontenc}
\title{Sustained changes to urban mobility after COVID-19 amplified socio-economic inequalities in Latin America}

\author[1,2,*,\textsuperscript{\textdagger}]{Carmen Cabrera}
\author[3]{Miguel González-Leonardo}
\author[1]{Andrea Nasuto}
\author[2]{Ruth Neville}
\author[1,\textsuperscript{\textdagger}]{Francisco Rowe}
\affil[1]{Geographic Data Science Lab, Department of Geography and Planning, University of Liverpool, Liverpool, UK}
\affil[2]{Centre for Advanced Spatial Analysis, University College London, London, UK}
\affil[3]{Centre for Demographic Urban and Environmental Studies, El Colegio de México, México City, México}

\affil[*]{Corresponding author: c.cabrera@liverpool.ac.uk}

\affil[ \textsuperscript{\textdagger}]{These authors have contributed equally to this study.}

\begin{abstract}
Urban mobility is central to economic activity, social inclusion and access to essential services. COVID-19 caused disruptions to mobility globally, yet its long-term impacts in less developed countries remain poorly understood. Using over 170 million anonymised mobile phone records from Meta-Facebook users in Argentina, Chile, and Colombia (March 2020–May 2022), we find sustained changes in mobility across socioeconomic and rural–urban gradients. We reveal that mobility recorded the sharpest declines and remained below pre-pandemic levels in most high-density and low socio-economic deprivation areas, while low-density and more deprived communities returned to baseline. These differences reflect the scale of the initial mobility shock, rather than subsequent recovery rates. Net mobility to urban cores remained consistently below pre-pandemic levels, suggesting a shift in their functional role. By revealing how COVID-19 reinforced mobility-related inequalities, we contribute novel evidence for planners and policymakers seeking to build more inclusive and resilient mobility systems.

\end{abstract}

\begin{document}

\flushbottom
\maketitle
\thispagestyle{empty}

Cities were epicentres of the COVID-19 pandemic. By November 2020, approximately 95\% of all the reported infections and fatalities had concentrated in a few large cities \cite{Pomeroy2021}. Stringency measures to contain the spread of COVID-19 reshaped the global patterns of urban mobility in the early months of 2020 \cite{nouvellet2021}. Social distancing, lockdowns, and business and school closures forced sharp declines in human mobility, notably in large cities \cite{florida2021}, responding to a need for limited social face-to-face contact and a transition to increased digital interaction. Against this backdrop, media headlines portrayed a gloomy prospect for cities pointing to an ``urban exodus'' with anecdotal evidence showing an intensification in the relative number of people moving from large cities to suburbs and rural locations \cite{paybarah2020new, marsh2020escape}.

Subsequent scientific studies examined these patterns, providing empirical evidence for widespread reductions in the intensity of both short- and long-distance movements from cities \cite{de2022overview}, particularly during the outbreak of the COVID-19 pandemic in the early months of 2020. While the extent of these changes varied widely across cities \cite{hunter2024city, kephart2021effect}, they were consistently documented across many developed and developing countries during the early months of 2020 \cite{nouvellet2021, rowe2023}. Mobility from dense metropolitan areas contributed to reinvigorating ongoing processes of counter-urbanisation and suburbanisation in some countries \cite{rowe2023}, or setting these processes in motion in others \cite{halfacree2024counterurbanisation}. In the United States, COVID-19 is attributed with a hollowing out of large cities, creating a ``doughnut'' shape in metropolitan areas that reflects movements away from core city areas to suburban and outer parts \cite{ramani2024working}.

Layered onto these patterns, existing evidence indicates that the impact of COVID-19 on human mobility was unevenly across socio-economic groups \cite{duenas2021changes, do2021association, elejalde2024}. Stringency measures to curb the spread of COVID-19 infections sought to restrict mobility, with exceptions made for those with essential occupations in healthcare, transportation and retail services, such as in supermarkets \cite{florida2021}. To a large extent, these occupations included lower-paid jobs and relied heavily in public-facing face-to-face interactions \cite{florida2021}. As such, less socio-economically deprived communities recorded larger reductions in mobility than more deprived groups, as the former could shift to remote working and thus avoid the need for commuting to work \cite{nathan2020}. 

While existing work has advanced our understanding of the COVID-19 impacts on human mobility in more developed countries \cite{gonzález-leonardo2022b, roweCalafiore2022, rowe2023, wang2022}, less is known about the extent and persistence of mobility changes in less developed nations beyond the year 2020. %More affluent groups are likely to experience the long-term benefits of hybrid working -- such as improved work-life balance, geographic flexibility and cost savings -- as they have greater access to digital technology, and have occupations which can be done remotely or partly remotely.
The lack of suitable data has been a major limitation to capture national-scale changes to human mobility in less developed countries during the pandemic \cite{ RoweCCA24}. Traditionally, census and survey data comprise the main source to study human mobility patterns in these countries \cite{bell2014}. Yet, these data systems are not frequently updated and suffer from slow releases \cite{rowe2023big, Cabrera-Arnau2023}, with census data for example being collected over intervals of ten years in most Latin American countries, and released with a considerable time lag \cite{bell2014}. Data resulting from social interactions on digital platforms have emerged as a unique source of information to capture human population movement in less developed countries \cite{rowe2023big}. Particularly, location data from mobile phone applications have become a prominent source to sense patterns of human mobility at higher geographical and temporal resolution in real time \cite{calafiore2023, iradukunda2025}.

Drawing on a large dataset of 170 million observations from Meta-Facebook users' mobile location data, we aim to assess the extent and pace of urban mobility recovery in Argentina, Chile and Colombia over a 26-month period from March 2020 to May 2022 following abrupt declines during early phases of the COVID-19 pandemic. Specifically, we seek to address the following set of questions:

$\bullet$ To what extent have declines in human mobility endured the COVID-19 outbreak? 
 
$\bullet$ How has the recovery of mobility levels varied across rural-urban and socioeconomic gradients?

$\bullet$ How has COVID-19 impacted human mobility in high-density urban cores?

Latin America provides an ideal test-bed for addressing these questions because of its exceptionally high levels of inequalities \cite{de2004inequality, carranza2023wealth} and urbanisation \cite{united2023world}. Half of the 20 most unequal countries on the planet are in this region. The average income Gini index of the region is 4 percentage points higher than Africa's and 11 higher than China's \cite{milanovic2016global}. Cities display some of the starkest inequalities \cite{habitat2022world}. Reportedly, their public transit riderships have remained well below pre-pandemic levels \cite{bloomberg2025transit}. Currently, over 80\% of the population in Latin America live in urban areas. By 2050, this share is predicted to reach 89\% \cite{habitat2022world}. Developing an understanding of human mobility in Latin America is thus important to support sustainable and inclusive cities \cite{habitat2022world}.

\section*{Results}

\begin{figure}[h!]
\centering
\includegraphics[width=\linewidth]{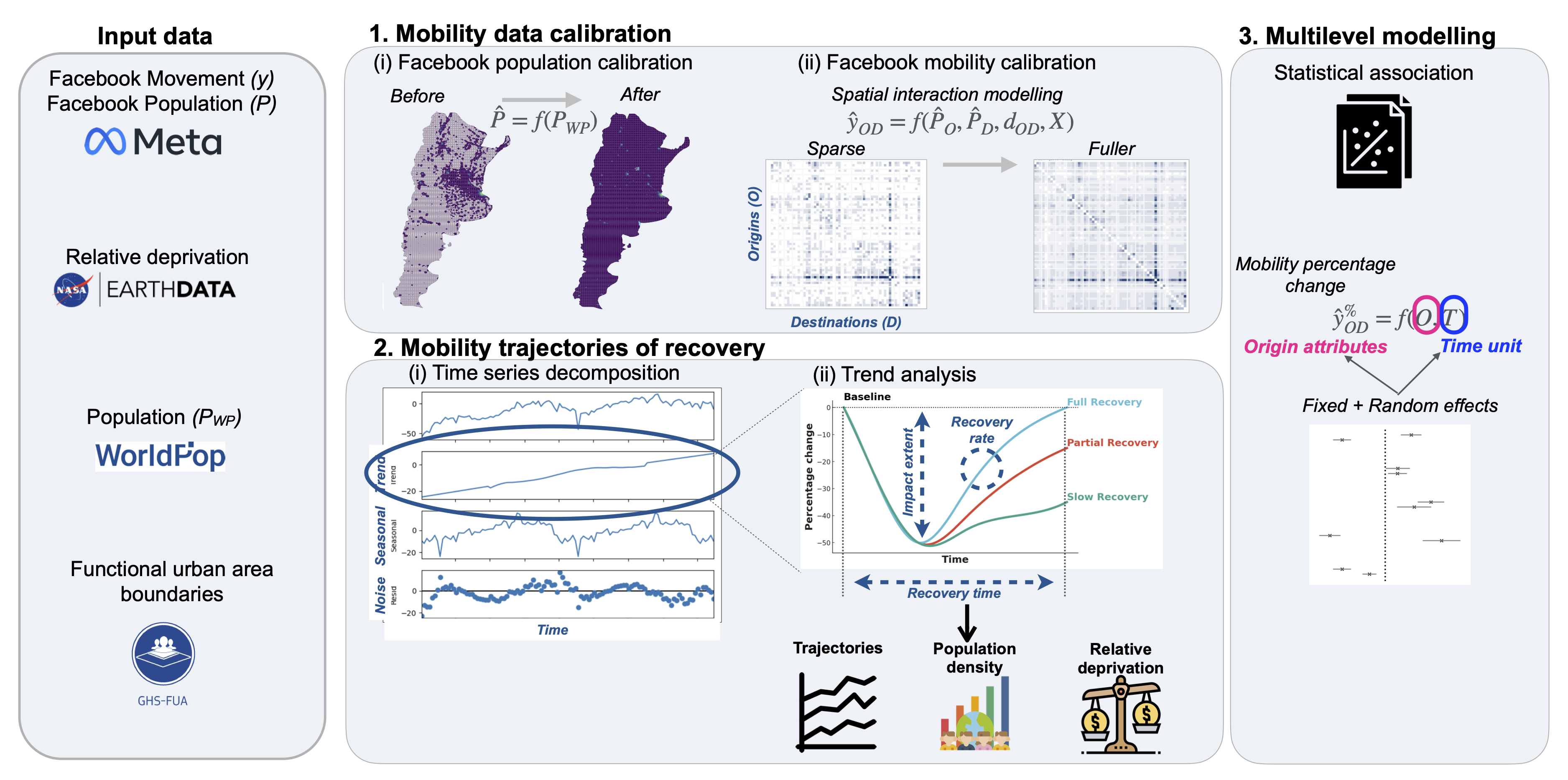}
\caption{Overview of data input, calibration and analysis.}
\label{fig-summary}
\end{figure}

We analysed post-COVID-19 mobility patterns over time relative to pre-pandemic baseline levels, using anonymised location data from Meta-Facebook users in Argentina, Chile, and Colombia, combined with high-resolution geospatial datasets (see Figure \ref{fig-summary}). Details about our approach can be found in the Methods section. We focused on daily mobility flows as a marker of urban vibrancy and economic activity, and assessed differences in mobility patterns across rural-urban and socio-economic deprivation gradients. We defined five population density classes based on Worldpop data and three socioeconomic deprivation deciles based on NASA's Relative Deprivation Index (RDI). The resulting population density classification mimics the degree of urbanisation definition from the Global Human Settlement Layer (GHSL) \cite{pesaresi2016ghsl}, representing areas along the rural-urban continuum: the highest density areas characterise urban centres, such as Central Business Districts (CBDs); followed by dense urban areas; semi-dense and suburban areas; rural areas; and, low-density rural areas. The resulting deprivation categories represent low, medium and high deprivation areas. We assessed the effects of population density and socio-economic deprivation separately, and do not include interaction effects, as the joint distribution of these variables showed limited overlap across several category combinations (see Supplementary Figure (SF) 8).

We addressed important sources of common biases in mobile phone app data for our analysis. As most digital trace data, mobile phone data do not necessarily provide a representative picture of the overall population, largely because of algorithmic decisions and changes in users' engagement over time \cite{rowe2023big, barreras2024exciting}. We know that Facebook suppresses population counts below 10 to protect users' privacy and that user engagement declined over time in our dataset. To mitigate these issues, we applied a series of adjustments (see Figure \ref{fig-summary}). First, we modelled the Facebook user population data as a function of Worldpop population data to generate a complete spatial population surface. Second, we modelled the Facebook mobility data to estimate origin-destination flows that were suppressed by the Facebook algorithm. Third, we used the temporal median of the sum of active user counts divided by the sum of active user counts across all spatial units on a given day to normalise for fluctuations in Facebook user engagement over time. A full description of these procedures is provided in the Methods Section. 

\subsection*{COVID-19 led to lasting disparities in mobility patterns}

\begin{figure}[h!]
\centering
\includegraphics[width=\linewidth]{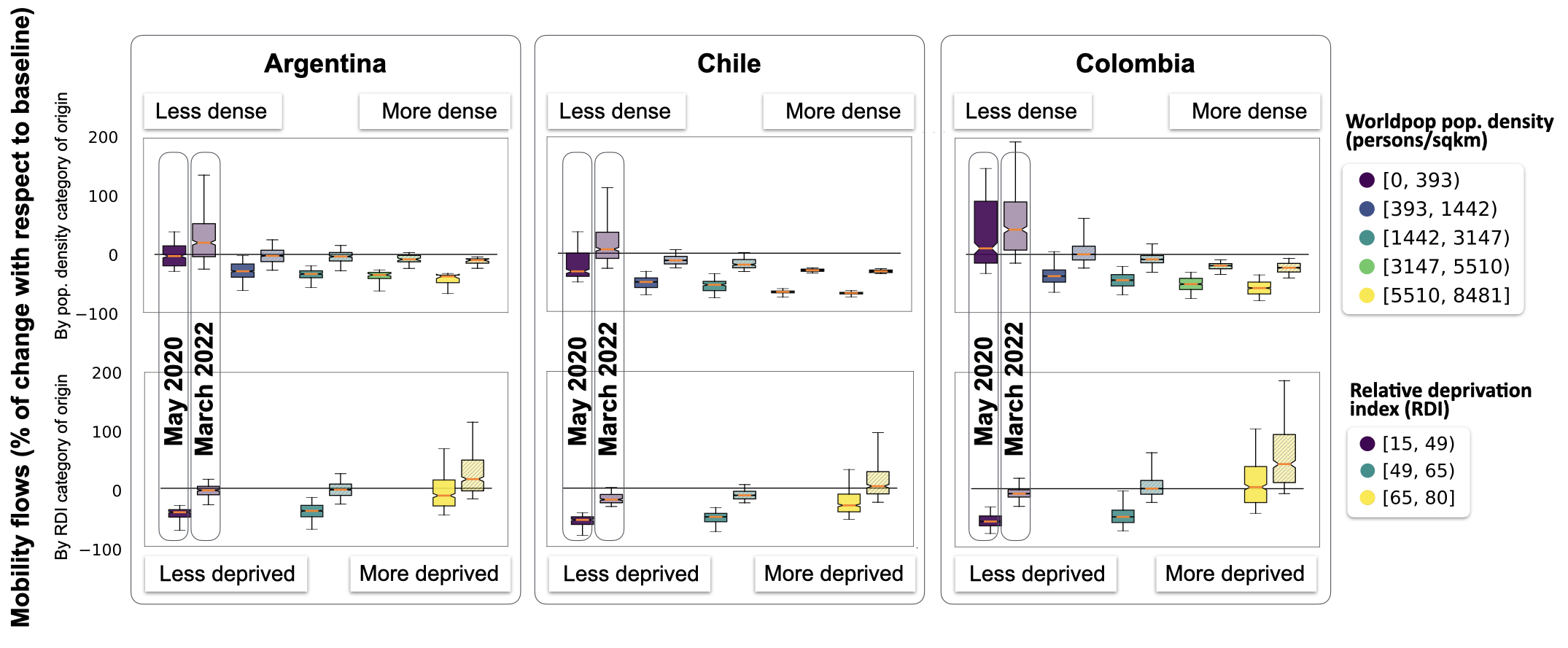}
\caption{Percentage change in mobility counts in May 2020 and March 2022, relative to the pre-pandemic baseline levels, by population density category and deprivation index (RDI) category, Argentina, Chile and Colombia.}
\label{fig-impact}
\end{figure}

We first compared the change relative to baseline levels of mobility during the early in the pandemic in 2020, and once most restrictions were lifted in 2022, as shown in Figure \ref{fig-impact}. Boxplots display the percentage change distribution of areas based on population density and relative deprivation level categories, with a $y = 0$ line representing baseline levels. Scores above this line indicate increased mobility from pre-pandemic levels, while values below represent a reduction. 

The results consistently show larger and more persistent declines in mobility in high-density areas compared to low-density areas. This persistence of reduced mobility levels in high-density areas is a novel finding, suggesting that the COVID-19 pandemic led to significant behavioural mobility changes in cities which have endured over time. The consistency of such a pattern is remarkable taking place in all three Latin American countries in our sample, showing little discrepancy within countries (as shown by relatively narrow box plots). In contrast, low-density areas show mobility patterns closer to pre-pandemic levels and greater variability. The large variability is due to the small population bases in these areas, where minor fluctuations in movement can result in disproportionately large percentage changes.

Our findings also evidenced enduring socio-economic disparities on the impact of the pandemic on mobility patterns. Prior research demonstrated that, early in the pandemic in 2020, a social gradient emerged, leading to sharper declines in mobility in less deprived areas compared to more deprived communities. We find that this social gradient effect has endured over time, as workers from more deprived areas are likely travelling as much (or more) as they used to prior to the outbreak of the COVID-19 pandemic - indicated by box plots stretching above baseline levels. These patterns are remarkably consistent across all three countries. This is despite differences in the initial extent of mobility decline across countries in 2020. Relative to baseline levels, initial reductions in mobility were larger in Chile (45-50\%) than in Colombia (40-45\%) and Argentina (25-30\%).

% We generated time series of the percentage change in mobility flows relative to the baseline, covering the entire dataset period. Figure \ref{fig-recovery-combined} a) and b) shows this evolution, disaggregated by the population density and relative deprivation index category of the area where the outflow originates. Time is represented on the $x$-axis while the intensity of movement is depicted on the $y$-axis, measured as the percentage change in the number of outflows relative to pre-pandemic baseline levels. To provide context in the evolution of mobility levels over time, we colour the plot background using the stringency index, which quantifies changes in the level of COVID-19 non-pharmaceutical interventions, such as social distancing and lockdowns. The time series serve as input for the next stage of analysis, aimed at further characterising the disparities in the recovery of mobility.

% \begin{figure}[h!]
% \centering
% \includegraphics[width=0.8\linewidth]{figures/Fig-recovery-combined-design.png}
% \caption{Percentage change in outflows relative to pre-pandemic baseline levels in origin areas, for Argentina, Chile and Colombia, from mid-April 2020 to mid-April 2022: a) by population density category, b) by relative deprivation index (RDI) category.}
% \label{fig-recovery-combined}
% \end{figure}

\subsection*{Early mobility declines during the pandemic shape long-term levels of mobility}

\begin{figure}[H]
\centering
\includegraphics[width=\linewidth]{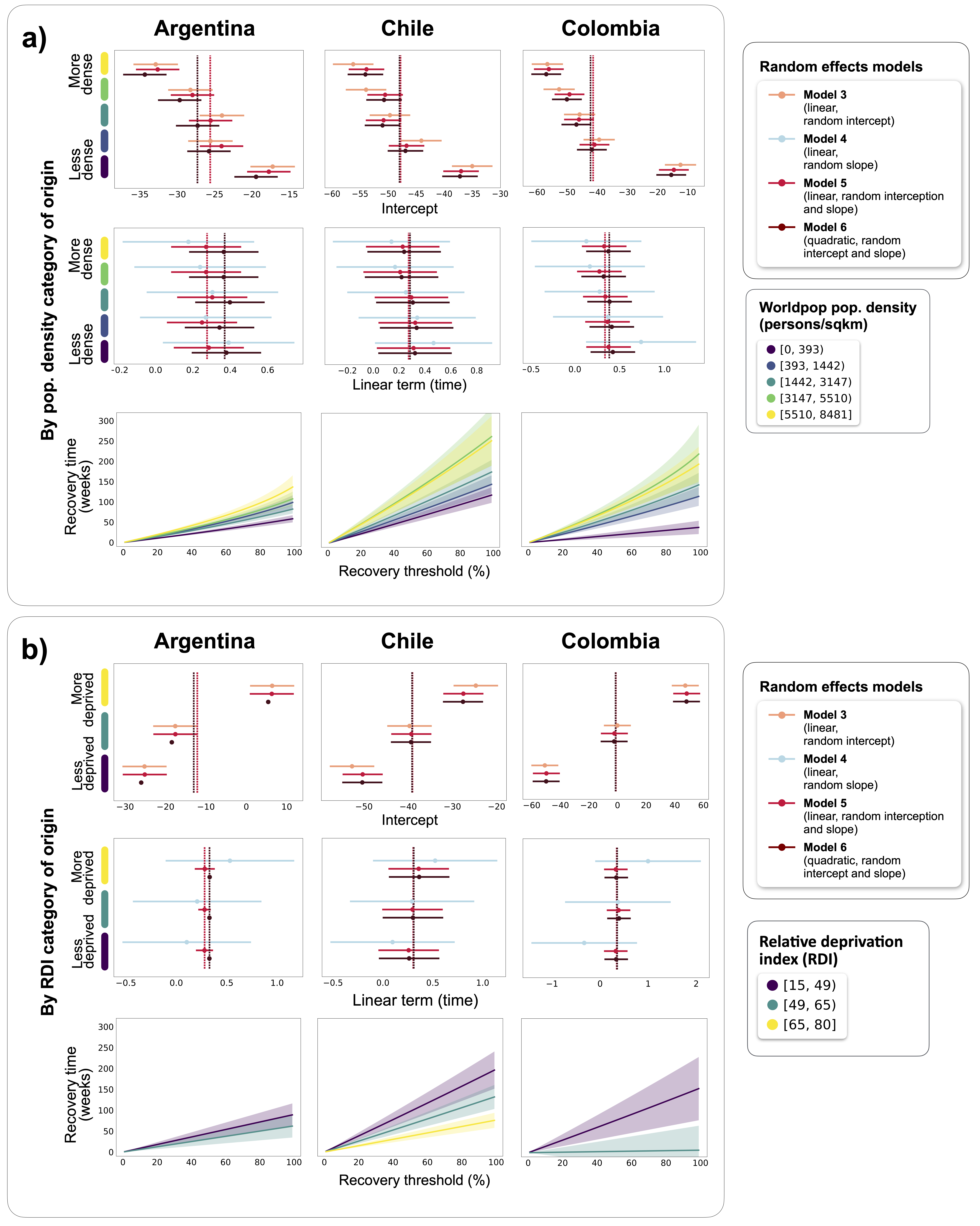}
\caption{First and second row of each panel represent mixed-effects modelling estimates (random intercept and random linear time term); bars denote 95\% confidence intervals; colours encode different model specifications; vertical dashed lines represent the estimated values for fixed effects of the corresponding terms (i.e. intercept, linear time). Third row represents estimated recovery times at different thresholds. Panel a) displays results by population density category and b) by relative deprivation category.}
\label{fig-re-combined}
\end{figure}

Discrepancies in the resilience of areas may underpin the observed differences in mobility levels. To quantify the resilience of areas, we model the trends of mobility recovery to pre-pandemic levels. To this end, we decomposed the time series of the percentage change in mobility into trend, seasonal, and residual components based the Seasonal and Trend (STL) decomposition using Locally Estimated Scatterplot Smoothing (LOESS) \cite{cleveland1990stl}. The time series are displayed in the SF9. We modelled the trend component as a function of time, stringency levels and attributes of the places of origin, including population density and relative deprivation, in a mixed-effects modelling framework – see Methods section and Supplementary Table (ST) 2 for details. Fourteen different model specifications were fitted separately for each country. The full set of regression estimates are provided in ST3-8. Our stringency and time variables exhibited a strong negative correlation (Pearson correlation coefficient of -0.91, $p$-value $< 0.01$). Due to this high collinearity, the first and second rows of Figure \ref{fig-re-combined} (a) and (b) report parameter estimates from models that include only time as an explanatory variable (rather than both time and stringency). This decision is supported by the Akaike Information Criterion (AIC), which generally indicates better model fit (i.e. lower AIC scores) for time-only specifications. Among all models, Model 6 generally yields the lowest AIC score, indicating the best overall fit.

The model estimates reported in Figure \ref{fig-re-combined} correspond to coefficients for intercept and time terms from Models 3-6 for population density (a) and deprivation (b). The intercept represents the estimated percentage decline in mobility during the week when Meta began producing mobility records, correcting for seasonality and noise. Negative scores indicate reductions in mobility levels, relative to the baseline period. The time term captures the linear component of the weekly rate of mobility recovery, relative to baseline levels. A time quadratic term was also included to capture variations in the pace of mobility recovery over time as reported in ST3-8 for Models 6 and 7. Model 7 includes random effects for the quadratic term. This adds complexity to our model, but it does not improve the estimation (higher AIC) and is often insignificant (p-value$>0.05$). Estimates from various model specifications evidence the consistency of our findings. Additionally, we also estimated mobility recovery times, reported in both panels (a) and (b); that is, the time taken (in weeks) for mobility to reach various thresholds based on pre-pandemic mobility levels. For low-deprivation groups in Argentina and Colombia, recovery times are not reported, as average mobility levels for these groups did not fall below the baseline during the observation period. The Methods and Supplementary Information (SI, section 4) describe how the recovery time estimates and uncertainty bands were computed. ST21-24 report recovery time estimates by country at four different thresholds (25\%, 50\%, 75\% and 100\%), and by population density or RDI category.

First, our random intercept estimates is consistent with our evidence in Figure \ref{fig-impact} revealing a descending gradient across rural-urban and socioeconomic structures during early stages of the pandemic. Mobility declined the sharpest in highest-density and least-deprived areas, while low-density population and more deprived areas registered small declines. These estimates suggest that populations in affluent and more dense urban areas have more
flexibility to adapt to early COVID-19 restrictions by reducing their mobility levels and turning to remote work. Poorer and less populated communities had less flexibility to transition to remote work.

Second, our regression estimates show differences in random intercepts and overlapping confidence intervals for our random time slopes. These results suggest that differences in mobility recovery across rural-urban and socio-economic gradients over time reflect differences in the magnitude of the initial mobility shock (intercepts), rather than in recovery rates (time slopes) following the initial COVID-19 outbreak. That is, the magnitude of the initial mobility shock tended to shape mobility recovery differences across rural-urban and socio-economic gradients.

indicate no statistically significant difference in the rate of mobility recovery across rural-urban and socio-economic gradients. Yet, recovery time estimates indicate a consistent grading across rural-urban and socio-economic structures. Higher density and less deprived areas display longer recovery times,
indicating that mobility is estimated to take longer to reach pre-pandemic mobility levels. Differences in recovery times along rural-urban and socio-economic gradients are driven by the magnitude of the initial mobility shock (captured by our intercept estimates), rather than by differences in recovery rates (captured by our time estimates).

Third, recovery time estimates (Figure \ref{fig-re-combined} bottom panel) indicate longer recovery periods in mobility for higher population density and less deprived areas. The lowest population density areas in Chile are estimated to take 132 weeks to achieve 100\% of observed pre-pandemic mobility levels, but only 201.87 weeks in the highest density areas. The differences in recovery times are consistent across countries. Yet, wide variability exists between them. Chile systematically displays the longest estimated recovery times. They tend to be twice as large as those recorded for Argentina and Colombia, where stringency measures implemented at the outset of COVID-19 triggered a rise in mobility in the low density areas, and highly deprived communities, exceeding pre-pandemic levels (SF9).

Overall, our findings indicate that the degree of persistence of the immediate shock of the COVID-19 pandemic on mobility
levels. They suggest that post-pandemic mobility disparities reflect the immediate shock of COVID-19, rather than differences in recovery rates across communities. COVID-19 appears to have led to new mobility regimes reflecting pre-existing urban and socio-economic inequalities. Our findings indicate that the unequal mobility declines observed in March 2020 across rural-urban and deprivation gradients persisted well beyond the lifting of restrictions and vaccine rollout, lasting at least two years after the onset of COVID-19.

\subsection*{Dense urban cores report persistent reductions in net mobility outcomes}

Next we examine the resilience of high-density urban core areas as attractors of daily mobility. Prior work identified a hollowing out of cities since the COVID-19 outbreak, or ``doughnut effect'', with a net loss of daily mobility into urban centres \cite{ramani2024working}. However, such evidence is limited outside the US context \cite{ramani2024working, gonzalez2023donut}. We analysed the evolution of netflows (i.e. inflows minus outflows) in relation to pre-pandemic levels. We measured netflows for all urban cores, or CBDs in the country, and only for the capital's urban core (i.e. Buenos Aires, Santiago and Bogotá), to assess the extent to which flows to the capital may influence the results. We used Functional Urban Areas (FUAs) boundaries from the GHSL to define urban regions as they delineate daily commuting activity spaces \cite{ceu.jrc.2019}, and for each urban region, we distinguish the urban core as the most densely populated spatial unit of analysis (the densest Bing tile within the FUA, see Methods), which effectively contains the CBD. %In Buenos Aires, only one of our Bing tiles is classified as a very high-density area (shown in yellow); hence, no exchanges are recorded between very high-density areas within the metropolitan area of Buenos Aires, and thus no reported in the first panel of 

\begin{figure}[h!]
\centering
\includegraphics[width=\linewidth]{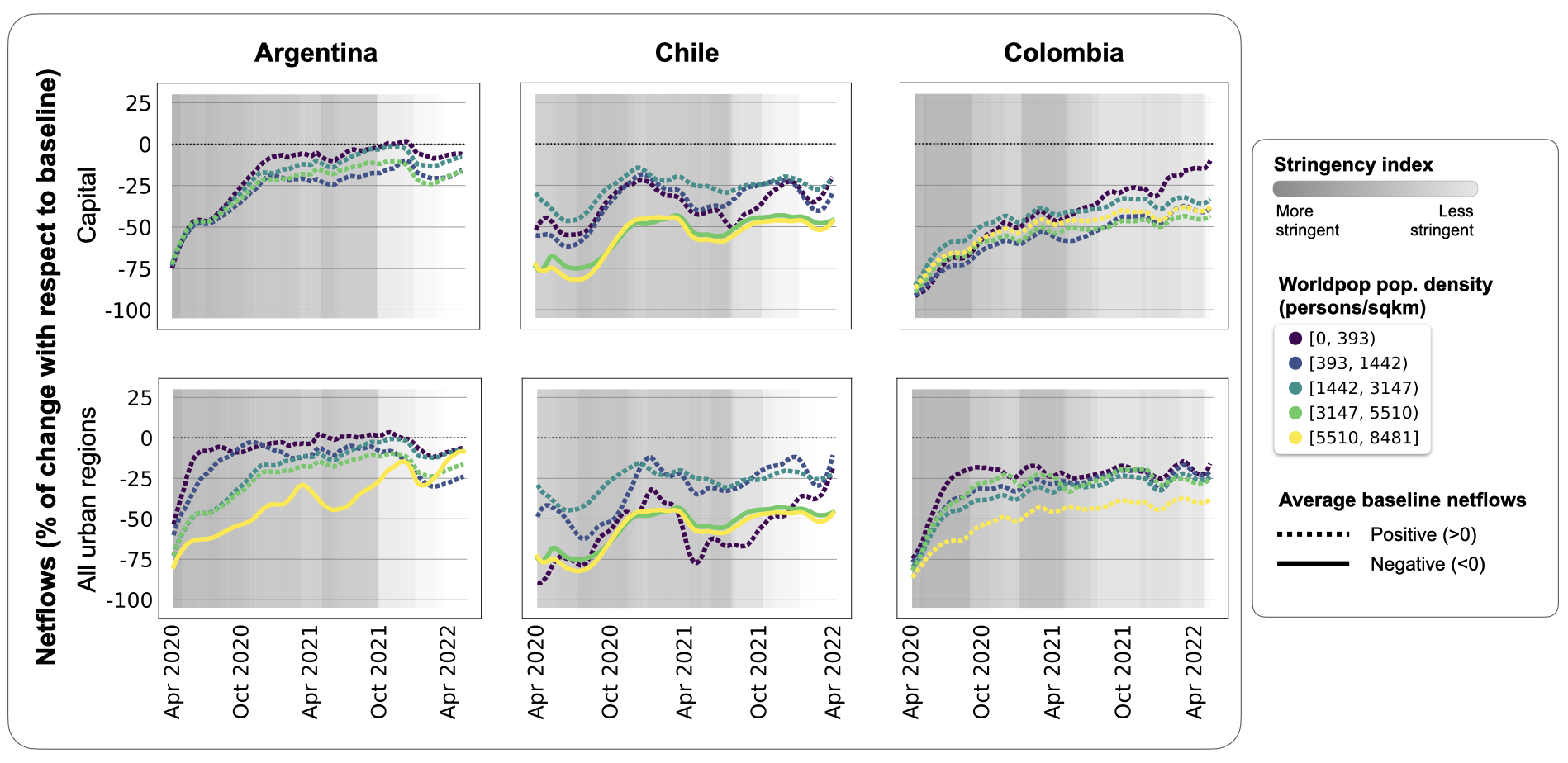}
\caption{Evolution of netflows for urban cores corresponding to the capital urban region and all urban regions (defined as Functional Urban Areas), by population density category for Argentina, Chile and Colombia, April 2020 to May 2022. Netflows are expressed relative to pre-pandemic baseline levels. Dashed and solid lines indicate, respectively, positive or negative pre-pandemic netflows on average.}
\label{fig-fuasflows}
\end{figure}

Figure \ref{fig-fuasflows} displays the evolution of these netflows by population density category. We seek to assess the extent of loss population in urban cores due to mobility and its durability since the onset of COVID-19. We evaluate if urban cores in Latin American countries recorded declines due to mobility redistributing people to moderate population density areas (suburbanisation); to low-density areas (counter-urbanisation); or both. In Argentina, only the area corresponding to the urban core in Buenos Aires is classified in the highest density category, so no mobility interactions with other high-density areas are captured when analysing just the capital's urban core; hence, no yellow curve for the first panel in Figure \ref{fig-fuasflows} is reported. We used a mixed-effects modelling framework to analyse the time series in Figure \ref{fig-fuasflows} and estimated recovery times at multiple thresholds (see ST9-20 and ST23-24). In some areas, netflows to urban cores did not recover. As a result, recovery times could not be estimated for these areas (ST9-20 and ST23-24).

%When the baseline netflow is positive, a negative percentage change between -100\% and 0\% indicates that the netflow decreases compared to the baseline but remains positive. Conversely, when the baseline netflow is negative, a percentage change between -100\% and 0\% suggests that the netflow remained negative during the crisis, though its absolute value was smaller than in the baseline period.

Our findings suggest a prevalent pattern of reduced netflows for urban cores following the enforcement of COVID-19 restrictions. Dashed lines indicate that netflows in urban cores incoming from other areas were positive pre-pandemic. However, the negative relative change since the COVID-19 outbreak reveals a reduction in the magnitude of these netflows. This reduction has predominantly become less acute over time across the rural-urban hierarchy in all three countries, but it remained below pre-pandemic levels following the lifting of COVID-19 restrictions in May 2022. These patterns reflect that while mobility inflows into core areas continued to exceed outflows, they shrank resulting in a smaller net balance. In Argentina and Chile, particularly in the capital city of Buenos Aires and Santiago, patterns of negative netflows are consistent with systematic evidence of net population losses due to internal migration in these metropolitan centres over the last two decades \cite{rowe2013spatial, vignoli2019efectos, rodriguez2022migracion}.

%We model the trend component of the time series in \ref{fig-fuasflows}. Assuming the quadratic model in equation \eqref{model-trend}, we estimate recovery times for netflows according to equation \eqref{recovery-time-formula} (all values reported in Supplementary Tables 22-23). We observe that in many cases, mobility never reaches baseline levels, particularly in urban areas from Chile. Netflows between the capital's urban core and the less densely populated areas within the surrounding urban region tend to return to baseline levels more quickly. This is likely due to the fact that residents in these areas rely more on high-density urban cores for essential daily activities.

%The patterns observed in high-density urban cores across all urban regions are similar to those found in capital cities. However, netflows between urban cores and less densely populated areas within urban regions tend to rebound more quickly toward pre-pandemic levels than when considering only the capital. This initial rebound is followed by a stabilisation of the percentage change, and netflows seem to never return to normal. This pattern further suggests that urban cores have lost some of their appeal, even among nearby low-density areas that may have previously been more dependent on them.

\section*{Discussion}

% I am getting a sense that we are saying that mobility is a sign of vulnerability: low-density and highest deprived areas - lower GDP country display closer mobility levels to prepandemic and faster recovery times. Would be good to look for literature on this. I can think of how mobility levels evolve with development, and there is a stage, they do grow with development before decreasing as areas become more developed.

Using Meta-Facebook origin-destination mobility data derived from GPS mobile phone locations, we assessed the durability of changes in mobility patterns across rural–urban and socio-economic gradients in Latin American countries during the beginning of the pandemic outbreak in March 2020 and following the rollout of vaccines and relaxation of lockdown measures in May 2022. Prior work showed that the pandemic resulted in large reductions in mobility levels in high-density urban areas and least deprived communities, reflecting reductions in commercial activity in urban areas, and a higher representation of affluent communities in jobs with higher potential for remote working \cite{bartik2020jobs}. Yet, existing evidence has focused on the immediate impacts of COVID-19 during early 2020 and on the restructuring of mobility patterns in more developed countries. We contribute to expanding this work by investigating the evolving impacts of COVID-19 on mobility beyond their initial aftermath extending to the second year of the pandemic in less developed countries.

We find evidence of sustained reductions in mobility levels in high-density urban areas and among the least deprived communities. These declines, first observed in the early months of the COVID-19 pandemic, persisted two years later, despite widespread vaccine rollout and the lifting of lockdown measures. We also find a sustained reduction of net balance of mobility flows in core urban areas. Historically, cities are critical engines of innovation, economic growth and prosperity \cite{Glaeser10}, enabling the emergence of agglomeration economies and facilitating knowledge exchange \cite{storper2004}. Our findings suggest that some of these benefits may have diminished. COVID-19 accelerated the adoption of technologies, online services and remote working at an unprecedented scale, reducing the need for physical co-location and eroding the vibrancy of urban centres \cite{Alexander21}. As people engage in virtual activities, planning strategies should be reoriented to repurpose existing infrastructure and develop new required infrastructure, to accommodate the ways in which people move and use cities in a post-pandemic world. Transport systems, for example, may need to be redesigned to accommodate less frequent trips to core urban areas, and less frequent but long-distance commuting trips from less dense areas locations to employment centres. Digital infrastructure in more remote areas may require an upgrade to accommodate improved connectivity and seamless virtual data transfers.

Importantly, our findings reveal persistent differences in mobility recovery trends across rural-urban and socio-economic gradients. We find that disparities in the initial impact of the COVID-19 pandemic on mobility across population groups largely account for the sustained differences in recovery trajectories. Notably, while the magnitude of the initial shock varied significantly across rural-urban and socio-economic groups, the subsequent rates of mobility recovery have remained relatively uniform over the long term. Higher-density and less deprived areas display shorter overall recovery times compared to lower-density and more deprived areas. These disparities reflect differences in the nature of employment \cite{ILO21} and can arguably enhance pre-existing socioeconomic inequalities, as daily mobility entails financial and non-financial opportunity costs, such as transport fares and travel time, which may disproportionally burden the most disadvantaged communities. Pandemic-induced differences in mobility behaviour across deprivation groups may continue in the future if current remote work practices are maintained. However, they may be about to change following recent shifts in employers' attitudes towards remote working, such as the presidential actions of the US government, with mandates for more days in the office \cite{WhiteHouse2025Return, CIPD2025LMO}. Overall, our findings suggest that efforts to reduce mobility disparities across rural-urban and deprivation divides should prioritise mitigating the initial shock of pandemics, as recovery trajectories tend to progress at similar rates across groups once the immediate disruption has passed.

In Latin America, the implications of our findings are specially serious, as the sustained pandemic-induced changes that we observed in mobility may exacerbate existing large levels of inequality. Latin America is considered one of the most unequal regions in the world \cite{WIReport22}. The bottom 50\% holding only 10\% of the national wealth in the region will probably face a greater relative social burden induced by their limited capacity to work remotely. At the same time, the top 10\% holding 55\% of the national wealth may continue to travel less often to urban centres, reducing their footprint, commercial activity in these areas, and eventually their overall vibrancy. If intensified and sustained, these trends will amplify already visible patterns of urban centre decay in some large Latin American cities, as we saw in European cities during the 1970s and 1980s \cite{THORSTEN}.

\section*{Methods}

\subsection*{Data}

\subsubsection*{Facebook Movement During Crisis}

We used anonymised aggregate mobile phone location data from Facebook-Meta users to capture population movements for Argentina, Colombia and Chile, covering a 24-month period from April 2020 to March 2022.
We used the Facebook Movement During Crisis datasets created by Meta and accessed through their Data for Good Initiative (\href{https://dataforgood.facebook.com}{https://dataforgood.facebook.com}).
The data are built from Facebook app users who have the location services setting turned on on their smartphone.
Prior to releasing the datasets, Meta ensures privacy and anonymity by removing personal information and applying three privacy-preserving processes \cite{Maas19}.
First, Meta adds a small, undisclosed amount of random noise to prevent the determination of precise counts for these areas.
Second, they apply spatial smoothing, using inverse distance-weighted averaging, to create a smoother data surface.
Third, they drop small counts to exclude records data counts below 10.

The Facebook Movement datasets provide information on the aggregated number of Facebook app users moving between pairs of locations from April 2020 to March 2022 during the COVID-19 crisis period.
The data is spatially aggregated into tiles according to the Bing Maps Tile System developed by Microsoft \cite{MicrosoftBingMaps}.
The Facebook Movement data are spatially aggregated into Bing tiles at resolution level 10 for Argentina and Chile, and level 12 for Colombia, which correspond to squares of 19.57 $\times$ 19.57 km, and 4.89 $\times$ 4.89 km, respectively. The data are temporally aggregated into three daily 8-hour time windows which are 00:00-08:00, 08:00-16:00 and 16:00-00:00, Pacific Time (PT).
These time windows translate to 05:00-13:00, 13:00-21:00, 21:00-05:00 for Argentina and Chile; and 03:00-11:00, 11:00-19:00, 19:00-03:00 for Colombia.
For each time window, the origin location of a user is defined as the most frequently visited location in the previous time window, while the destination location is the most frequently visited location in the relevant time window.

%We filtered the data to include only the morning time window for two reasons. First, this period is most likely to capture the majority of commuter activity. Second, focusing only on one part of the day avoids potential cancellation effects that could arise from considering all daily trips, as individuals typically engage in a round-trip pattern (e.g., travelling from home in the morning and returning in the evening). 

In addition, each dataset includes baseline movement counts, which represent the estimated number of people moving before COVID-19.
These baseline counts are calculated based on a 45-day period ending on March 10, 2020, using the average for the same time of day and day of the week within this period.
For example, the baseline for the 00:00–08:00 (PT) window on Mondays is calculated by averaging all Monday records for that time window across the 45-day period.
Further details about the baseline can be found in \cite{Maas19}.

Due to the small-count dropping privacy measure, movement counts during both the crisis and baseline periods are excluded if they fall below 10.
However, even when counts are not reported, the datasets consistently provide the percentage change in movement counts relative to the baseline.
This percentage change, denoted as $y^{\%}_{ijdt}$, is computed as follows \cite{Maas19}:

\begin{equation}
y^{\%}_{ijdt} = \dfrac{y^{c}_{ijdt} - y^b_{ijdt}}{y^b_{ijdt} - \varepsilon}\times 100.
\end{equation}

Here, $y^c_{ijdt}$ and $y^b_{ijdt}$ represent the crisis and baseline movement counts, respectively, between origin tile $i$ to destination tile $j$ on day $d$ and during time window $t$.
For the baseline values $y^b_{ijwt}$, the subindex $w$ denotes the day of the week corresponding to day $d$, as the baseline values are recorded by weekday, not by specific date.
A small constant $\varepsilon$, usually set to 1, is added to the denominator to avoid division by zero \cite{Maas19}.

\subsubsection*{Facebook Population During Crisis}

We also used the Facebook Population During Crisis datasets.
Like the Facebook Movement data, these datasets cover a 24-month period from April 2020 to March 2022 and provide information on the number of active Facebook users at specific locations during the crisis period. The Facebook Population data for Argentina, Chile and Colombia are spatially aggregated into Bing tiles at resolution level 11 for Argentina and Chile, and level 12 for Colombia, which correspond to 9.78 $\times$ 9.78 km and 4.89 $\times$ 4.89 km tiles respectively. The Facebook Population data is aggregated temporally into 8-hour time windows. A user's location is defined as the most frequently visited location within each 8-hour time window.
Similar to the Facebook Movements datasets, the Populations datasets include baseline counts, calculated by averaging values for the same day of the week and time of day over the 45-day pre-pandemic baseline period.

The same privacy-preserving techniques are applied to the Facebook Population datasets.
As a result, population counts during both the crisis and baseline periods are excluded if they fall below 10.
However, the datasets consistently report the percentage change in active user counts relative to the baseline.
For tile $i$, day $d$ and time window $t$, the relative change in active user counts $p^{\%}_{idt}$ is computed as:

\begin{equation}
   p^{\%}_{idt} = \dfrac{p^{c}_{idt} - p^b_{iwt}}{p^b_{iwt} - \varepsilon}\times 100, 
\end{equation}

where $p^c_{idt}$ and $p^b_{iwt}$ represent the crisis and baseline active user counts at tile $i$ on day $d$ and during time window $t$.
Once again, the subindex $w$ in baseline values $p^b_{iwt}$ denotes the day of the week corresponding to day $d$, and $\varepsilon$ is a small constant usually set to 1, added to the denominator to avoid division by zero \cite{Maas19}.

We use the term ``Facebook population'' in reference to the number of active Facebook app users.

\subsubsection*{Meta-Facebook data pre-processing}

We only included Facebook data for one time window per day: 08:00–16:00 Pacific Time (PT) corresponding to 13:00–21:00 in Argentina and Chile, and 11:00–19:00 in Colombia.
This time window is used to capture daily commutes.
Including data for the three time windows in the datasets could result in movement in opposite directions cancelling out and precluding us to capture any meaningful patterns.
To reflect our focus on one time window, we can drop the subindex $t$.

As a pre-processing step, we ensure that both the Facebook movement and population data are aligned in terms of spatial resolution for each country.
Since the Facebook Population datasets are available at a finer spatial resolution, we re-aggregate them to match the spatial resolution of the Facebook Movement data.

\subsubsection*{Worldpop population data}

We used data from Worldpop \cite{tatem2017} to classify our spatial units  by population density category, and  estimate missing baseline values in the Facebook population data.
We used gridded population data at a resolution of 1km$^2$ in raster format.
We performed a spatial join of the Facebook spatial units (Bing tiles) with the gridded population data and computed the sum of Worldpop populations corresponding to each of the Facebook spatial units.

\subsubsection*{Socioeconomic deprivation data} 

We used the Global Gridded Relative Deprivation Index (GRDI), Version 1 (GRDIv1) \cite{RDI2022} dataset as a measure of socioeconomic deprivation.
The GRDI data is available via NASA's Socioeconomic Data and Applications Centre (SEDAC) at a spatial resolution of 30 arc-seconds, or 1 km$^2$ approximately.
The index quantifies the relative levels of multidimensional deprivation and poverty, where a value of 100 represents the highest level of deprivation and a value of 0 the lowest.
We performed a spatial join of the Facebook spatial units and the gridded relative deprivation data. We computed the average RDI corresponding to each of the Facebook spatial units.

\begin{figure}[h!]
\centering
\includegraphics[width=\linewidth]{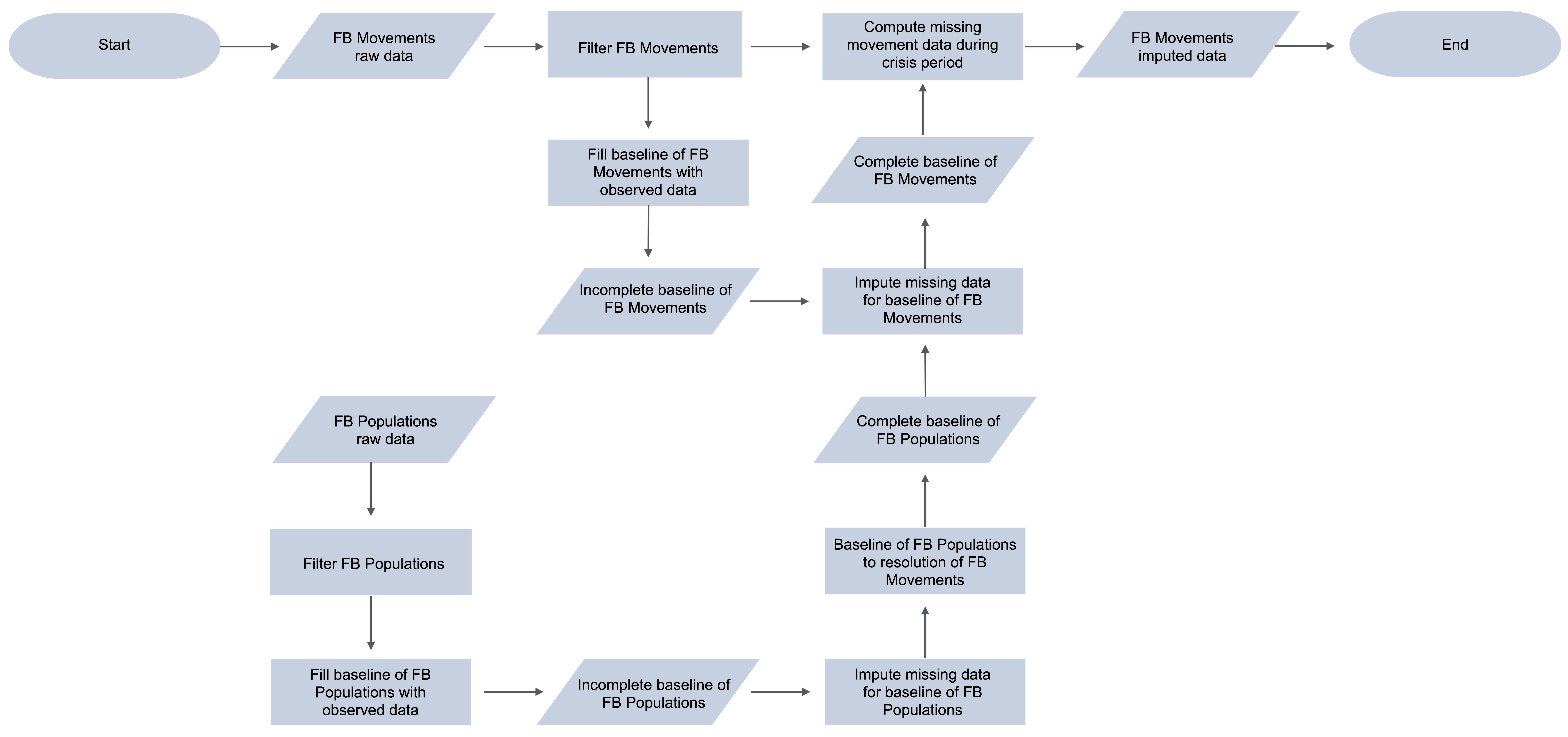}
\caption{Workflow diagram for data imputation method.}
\label{fig-dataimputation}
\end{figure}

A challenge in analysing the Facebook population counts and movements is the absence of records for small counts.
These records are removed to ensure privacy and anonymity. Yet, they may impact the analysis as the Facebook datasets include a high proportion of missing records, as shown in ST1, and missingness is not at random in these datasets. Missing values display a strong spatial pattern (see SF1). As such, sparsely populated spatial units, or small active Facebook user population could potentially be under-represented in the analyses. Simply removing the missing records from the analysis could lead to geographically biased results \cite{afghari2019}.
To address this, we designed a data processing method for missing data imputation. An overview of this data imputation method is provided in Figure \ref{fig-dataimputation}.

Additionally, we address a second challenge which is changes in the number of active Facebook users over time. To address this, we applied correction factors and time series smoothing to eliminate the fluctuations in the daily number of observations, ensuring that the representativeness of Facebook data across spatial units remains stable during the study period.
The steps of our data-processing workflow are described next.

\subsection*{Imputing Facebook Population data}

First, we identified all the baseline values reported in the Facebook Population datasets, which form the basis for creating a new dataset that includes all available baseline information.
Following \cite{duan2024}, we used linear models to estimate the missing Facebook population baseline counts based on Worldpop population data.
These models are fitted using ordinary least squares regression, with Worldpop data as the explanatory variable and the available baseline counts from the Facebook Population dataset as the dependent variable.
The estimation process and its accuracy are illustrated in SF2-4 for Argentina, Chile and Colombia, respectively.

We then used the complete baseline of Facebook population counts to compute missing Facebook population counts during the crisis period.
This is possible because, as mentioned above, Meta reports the percentage change in the number of counts with respect to the baseline $p^{\%}_{id}$, even if the counts are not reported due to numbers below 10.

\subsection*{Imputing Facebook Movement data}

The imputation of Facebook Movement baseline values is done through the use of spatial interaction modelling \cite{rowe2024spatial}.
We used baseline movement counts $y^b_{ijw}$ between pairs of origin $i$ and a destination $j$ tiles on weekday $w$, and modelled these counts as a function of the Facebook population count at the origin tile on the same weekday, the Facebook population count at the destination on the same weekday and the distance between origin and destination.
We also included indicator variables to capture the day of the week.
Mathematically, this model can be expressed as: 

\begin{equation}\label{eq:sim}
\mu^b_{ijw} = \beta_0 + \beta_1p^b_{iw} +\beta_2p^b_{jw} + \beta_3d_{ij} + \beta_4w + \varepsilon
\end{equation}

where: $\mu^b_{ijw} = E[y^b_{ijw}]$ is the expectation of the flow of people from tile $i$ to tile $j$ on the weekday $w$ during the baseline period; $\beta_0$ is an intercept, $p^b_{iw}$ and $p^b_{jw}$ are the Facebook population counts at the origin and destination on weekday $w$ during the baseline period, $d_{ij}$ is the distance between origin and destination, $w$ is a series of indicator variables capturing the day of the week, and $\beta_{0,1,2,3,4}$ are model parameters to be estimated from the observed data.
The error term is denoted by $\varepsilon$.
To estimate the model parameters, we used a Gaussian regression model, taking the log of the population at origin and at destination, and the log of the distance. SF 5 demonstrates the high correlation between the counts for observed flows and those predicted according to the spatial interaction model as specified in equation \eqref{eq:sim}.

SF6 provides additional validation of the spatial interaction model used to estimate missing baseline Facebook movement counts. This figure illustrates the Facebook population at the origin and destination of each flow during the baseline period, considering the imputed Facebook population baseline data. The results demonstrate that, for both observed and predicted data, high-volume flows typically occur between highly populated origin and destination locations, aligning with expected patterns.

We computed missing Facebook movement counts during the crisis period by considering the complete baseline of Facebook movement counts and the percentage change in the number of counts with respect to the baseline, which is reported in the Facebook Movement datasets even when the count is not reported due to its low value.

\subsection*{ Applying correction factors}

The total number of active users within a given time window shows daily fluctuations due to limitations in Internet connectivity and user data access options \cite{Maas19}.
Following \cite{yabe2020} and \cite{duan2024}, we applied a correction factor to mitigate the potential impact of these daily fluctuations on the results, since they could mask the mobility trends.
This approach assumes that the representativeness of Facebook data across spatial units remains consistent throughout the study period.

The adjusted number of movements between tile $i$ and tile $j$ on day $d$, $y'_{ijd}$ was obtained as:

\begin{equation}
y'_{ijd} = k_d \times y^c_{ijd}
\end{equation}

where: $y^c_{ijd}$ is the original number of movements between tiles $i$ and $j$ on day $d$ during the crisis period and $k_d$ is a correction factor computed as the median of the sum of active user counts across all days $d$ divided by the sum of active user counts across all spatial units on day $d$. Mathematically,

\begin{equation}
k_d = \dfrac{\mbox{med}_d\big(\sum_i{p^c_{id}}\big)}{\sum_i{p^c_{id}}}.
\end{equation}

\subsection*{Time series smoothing}

A time series for each pair of origin-destination tiles was generated using data processed as described above.
However, the resulting time series still contained missing values due to days when no data was reported for specific location pairs.
To ensure data completeness, we inputted missing values within the time series, replacing them with the average of the nearest 15 observations within the time series.
This number was chosen to provide an optimal balance between maintaining temporal proximity and ensuring sufficient data coverage.

Additionally, we applied a rolling-average smoothing technique to the resulting time series. This is done to reduce noise and enable the identication of underlying trends that might be obscured by short-term fluctuations.
This approach was used in the time series displayed in SF9.

\subsection*{Classification of tiles by urbanisation and socio-economic deprivation levels}

The geographic distribution of categories for each of the countries is displayed in SF7 for Argentina, Chile and Colombia, respectively. SF7 shows the proportion of the population across the various categories, based on both Facebook population counts and Worldpop population estimates. The Facebook population counts reflect the average number of active users across all weekdays in the pre-pandemic baseline period, after calibration.

In all three countries, the population distribution derived from Facebook data closely aligns with that of WorldPop. This shows that Facebook data reliably represent the population groups analysed in this study. Small discrepancies exist, for instance, in Argentina and Chile, where Worldpop data indicates a higher proportion of people living in low-density areas, suggesting that Facebook data under-represents populations in these regions. Similarly, in these countries, Facebook data shows an over-representation of the most affluent socioeconomic group. 

\subsection*{Trend analysis}

To quantify intergroup differences in the evolution towards pre-pandemic mobility patterns, we modelled the time series displayed in SF9. To this end, we extract the trend, seasonal and noise components of each time series using the `seasonal\_decompose()' method from the time-series models and methods API in Python's `statsmodels' package (version 0.14.4). We then modelled the trend component using seven linear mixed-effects model specifications, using`glmmTMB' library (version 1.1.10) in the R software package.

We estimated a series of linear and nonlinear mixed-effects models to analyse mobility trends during the COVID-19 pandemic. The general structure of the models is expressed as:

\begin{equation}\label{model-trend}
Y_{it} = \beta_0 + \beta_1 \cdot {t} + \beta_2 \cdot {t}^2 + \beta_3 \cdot S_{t} + \beta_4 \cdot C_i + b_{0i} + b_{1i} \cdot t + b_{2i} \cdot t^2 + \varepsilon_{it}
\end{equation}

where: $Y_{it}$ is the mobility count for origin $i$ at time $t$, $\beta$ terms represent fixed effects, $b_{0i}, b_{1i}, b_{2i}$ are random effects at the origin level (based on population density or RDI), $S_t$ is the stringency index at time $t$, $C_i$ is the category at the origin level, and $\varepsilon_{it}$ is the residual error. Not all components are included in every model. ST2 summarises the structure of each model.

Figures \ref{fig-re-combined} a) and b) illustrate the estimated random effects for Models 3--7. These models capture heterogeneity in both the initial decline and the rate of mobility recovery across areas along the population density or relative deprivation gradients. In some cases, standard deviations of the random effects are not reported due to non-convergence of the estimation algorithm. Full model estimates are provided in ST3, 5 and 7 (without stringency index) and 4, 6 and 8 (with stringency index) for Argentina, Chile, and Colombia, respectively.

\subsection*{Measuring recovery time}

We computed a measure of recovery times. This indicator quantifies the number of weeks required for mobility to return to specific recovery thresholds (e.g. 25\%, 50\%, 75\%, 100\%) relative to the baseline level. This measure is derived from the trend component of the relative change in mobility over time. Assuming a quadratic model for the trend component, the estimated recovery time reflects the pace at which mobility trends approach pre-disruption levels. We defined the recovery time at threshold $\alpha$ as:

\begin{equation}\label{recovery-time-formula}
    t_{\alpha} = \dfrac{-b + \sqrt{b^2-4c \alpha a}}{2c}
\end{equation}

where: $a$, $b$ and $c$ are the coefficients for the intercept, linear and quadratic terms in equation \eqref{model-trend} and $a$ and $b$ must satisfy $a<0$ and $b>0$. A derivation of equation \eqref{recovery-time-formula} can be found in the SI, section 4.1. We report estimates of $t_{\alpha}$ for all population density and deprivation categories using the estimates of coefficients $a$, $b$ and $c$ from Model 6 (or estimates of $a$ and $b$ from Model 5 in the case of Argentina when grouping by RDI, as Model 6 does not converge). Reported estimates for $t_{\alpha}$ are provided in ST21-23. Standard error have been computed as a function of the standard deviation of the estimates for $a$, $b$ and $c$ using error propagation. We included the formulas in the SI, section 4.2.

\bibliography{bibliography}

\section*{Acknowledgements}

The authors acknowledge the financial support of Research England through the Public Policy Quality-Related Pump-Priming Fund (project ID: 177990). CC and FR acknowledge the financial support of the UK's Economic and Social Research Council (project ID: ES/Y010787/1).

\section*{Author contributions statement}

C.C.: conceived the study, designed the methodology, carried out the data analysis, contributed towards the interpretation of results, wrote and revised the manuscript; F.R.: conceived the study, designed the methodology, contributed towards the interpretation of results, wrote and revised the manuscript; M. G.-L.: conceived the study, wrote and revised the manuscript; A. N. and R. N: carried out the data analysis, revised the manuscript. All authors gave their final approval for publication.

\section*{Additional information}

The code for this study will be made available on GitHub shortly.

%through the following link: \href{https://github.com/carmen-cabrera/latin-mobility-covid}{https://github.com/carmen-cabrera/latin-mobility-covid}.

\end{document}

\flushbottom

\maketitle


\section{Processing Facebook data}

\subsection{Missing values}

\begin{table}[h!]
\centering
\includegraphics[width=\linewidth]{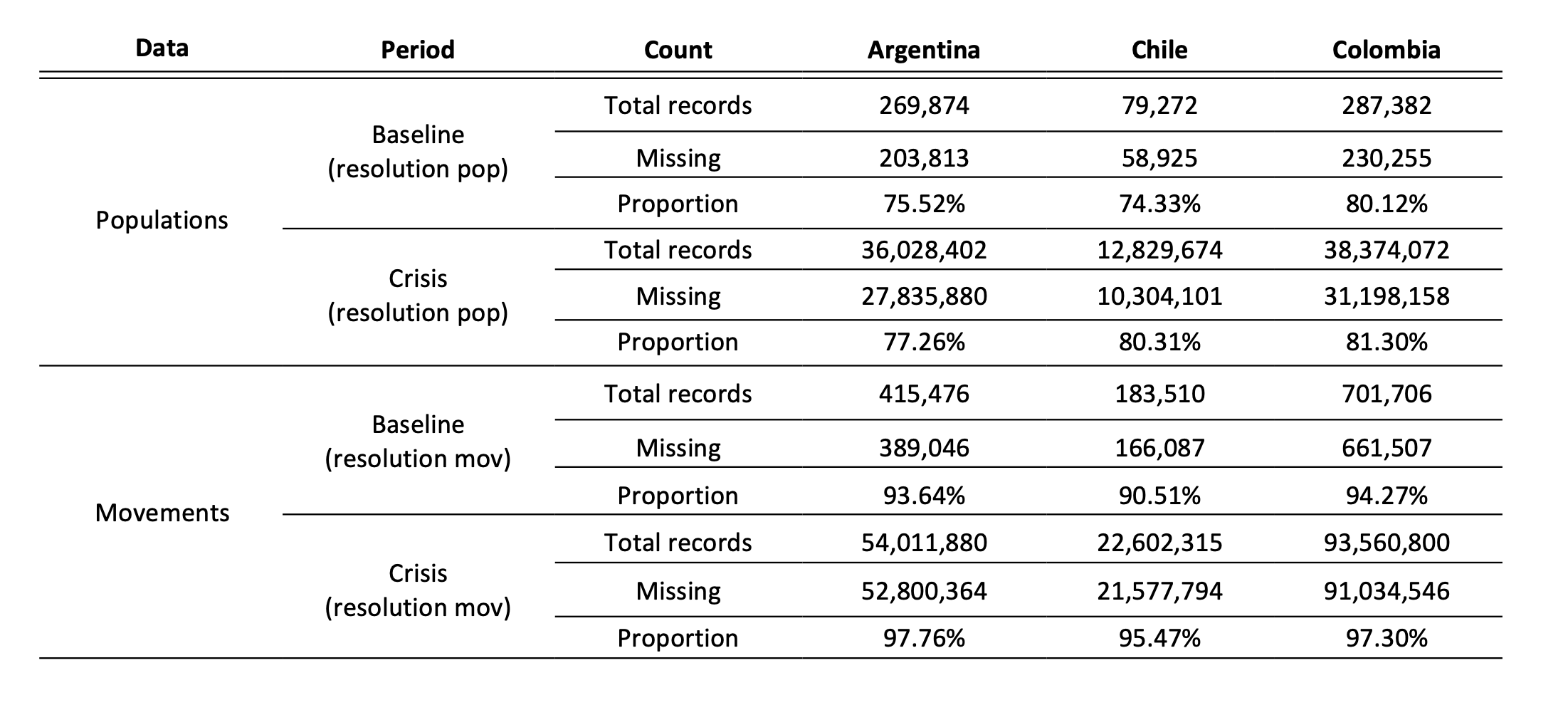}
\caption{Proportion of missing values for Facebook Population and Facebook Movements data, during baseline and crisis in Argentina, Chile and Colombia.}
\end{table}

\newpage

\begin{figure}[h!]
\centering
\includegraphics[width=.7\linewidth]{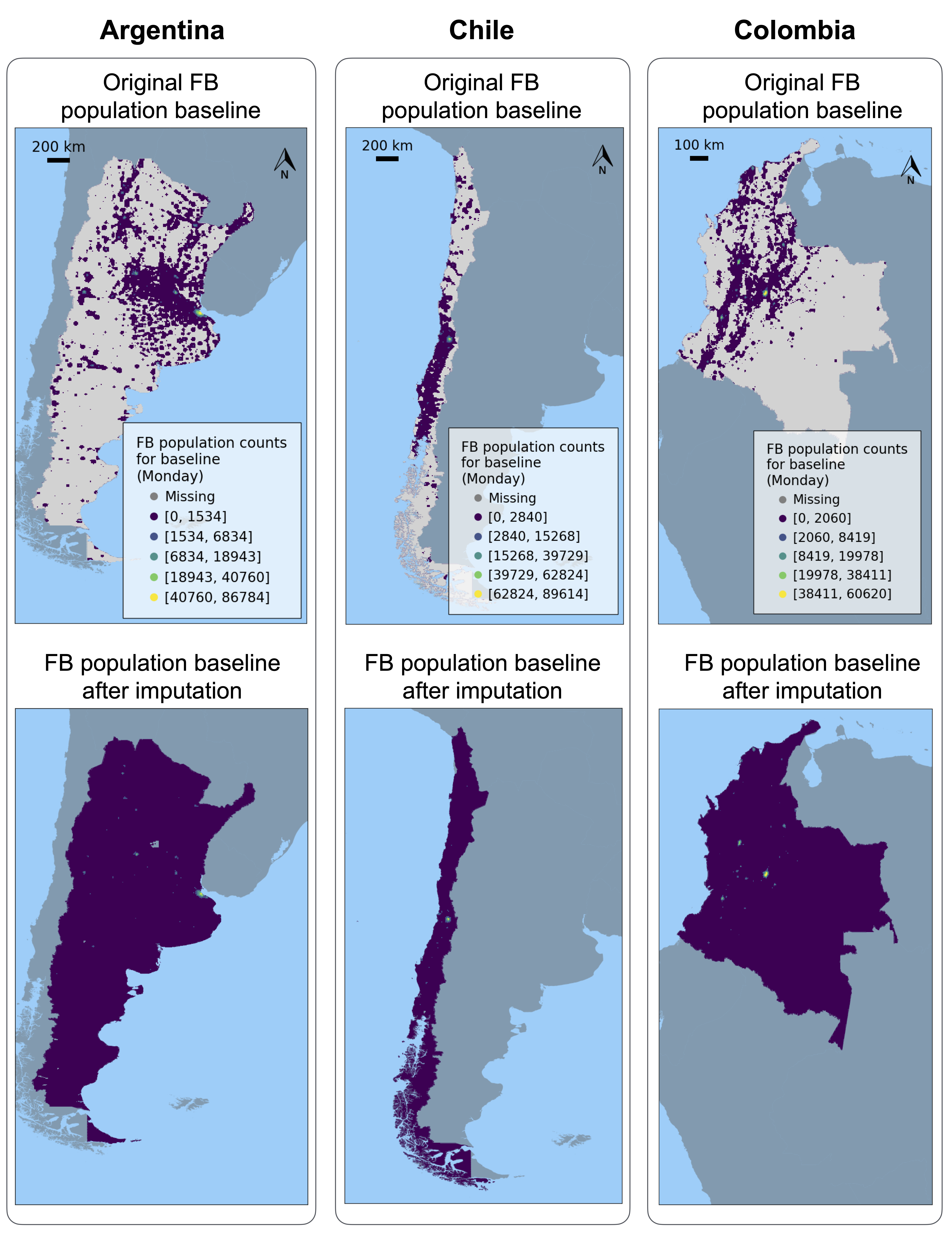}
\caption{Facebook Population baseline records corresponding to Monday, for Argentina, Chile and Colombia. The first row shows the original records, which include a high proportion of missing values. The second row shows the imputed baseline following a regression model with Worldpop data.}
\label{missing-baseline}
\end{figure}

\newpage

\subsection{Relationship between WorldPop population and Facebook Population baseline data for each day of the week}

\begin{figure}[h!]
\centering
\includegraphics[width=\linewidth]{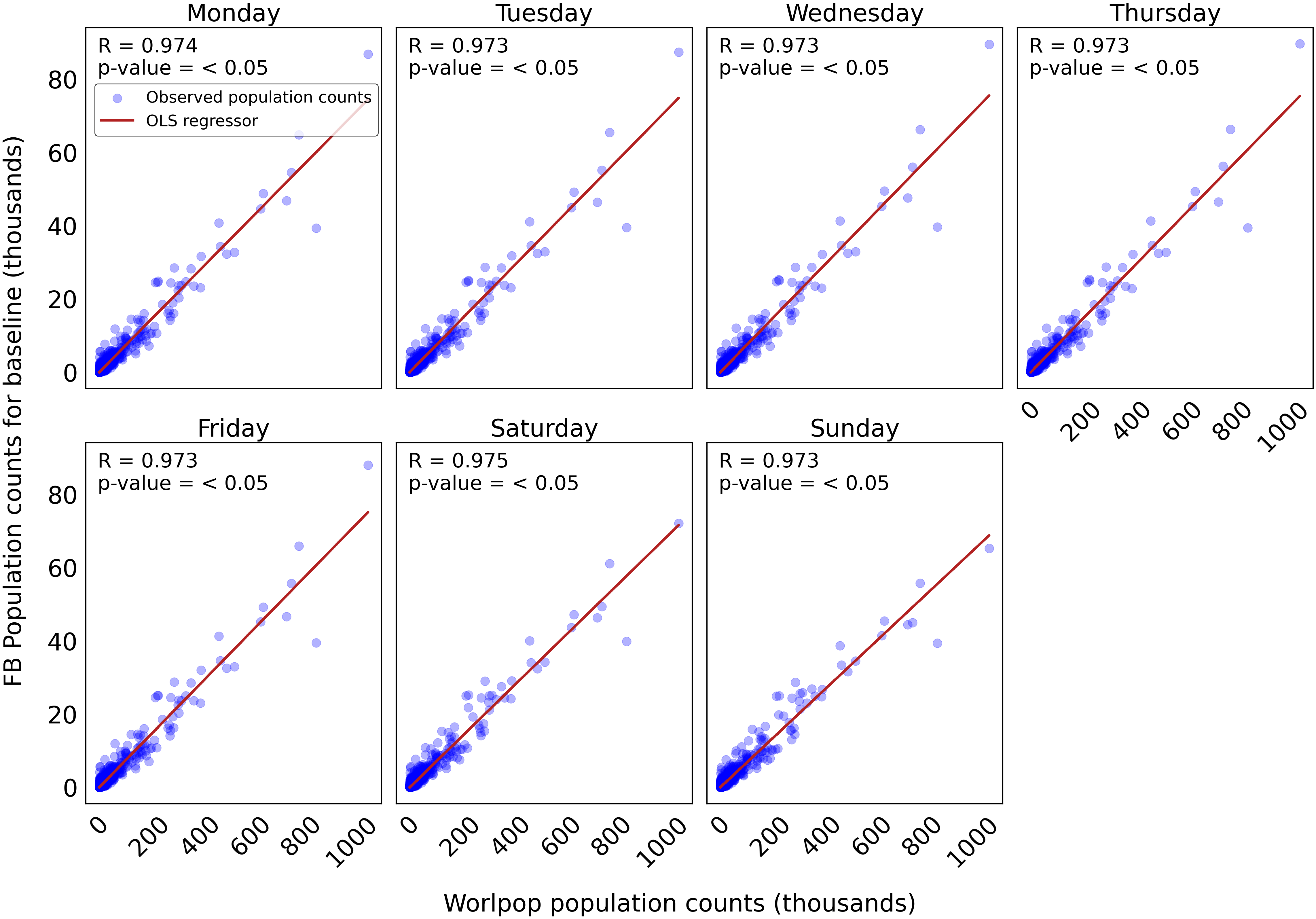}
\caption{Correlation between Worldpop population counts and Facebook Population counts during the baseline period in Argentina, for each day of the week.}
\label{correlation-wp-fbpop-ARG}
\end{figure}

\newpage

\begin{figure}[h!]
\centering
\includegraphics[width=\linewidth]{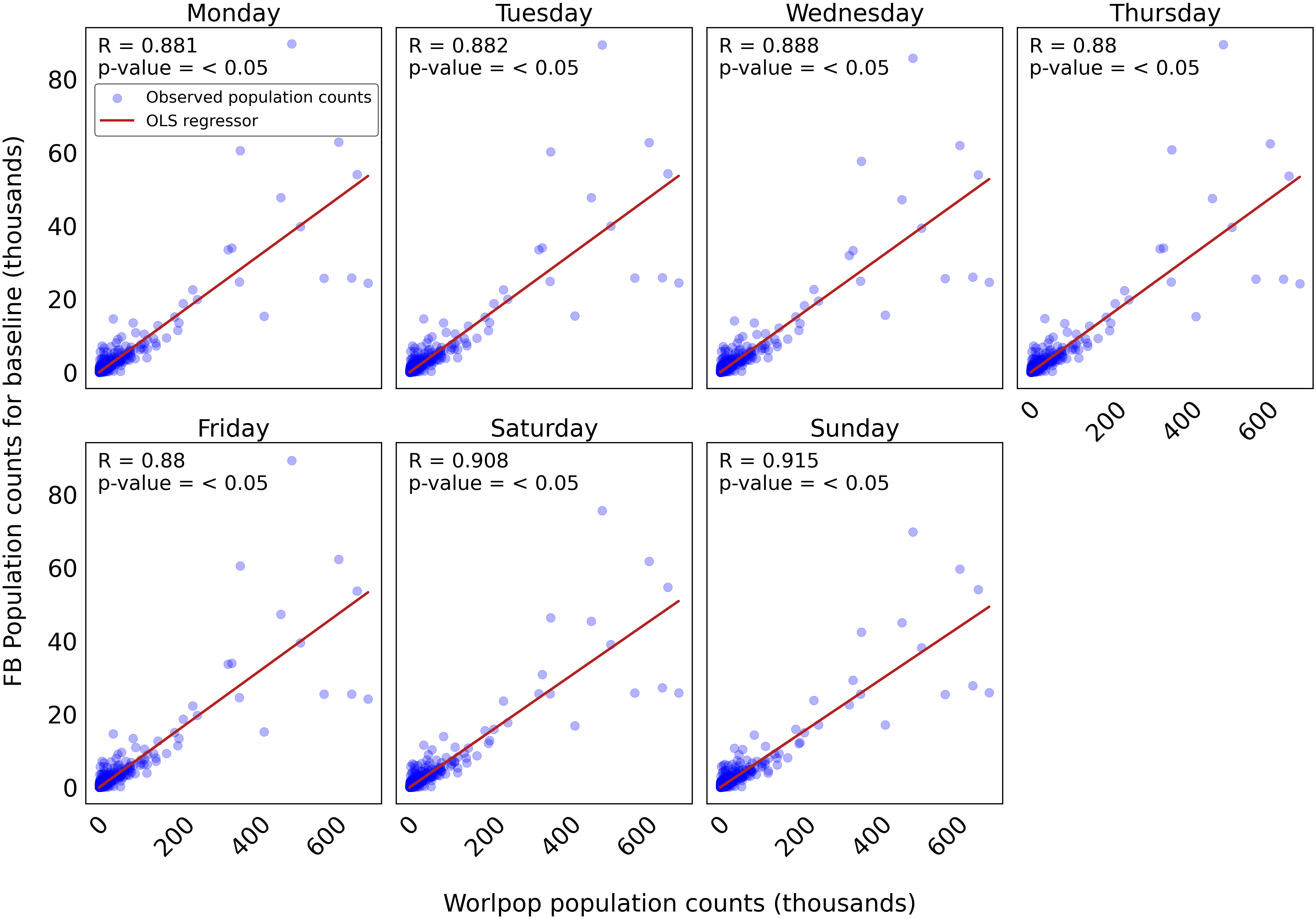}
\caption{Correlation between Worldpop population counts and Facebook Population counts during the baseline period in Chile, for each day of the week.}
\label{correlation-wp-fbpop-CHL}
\end{figure}

\newpage

\begin{figure}[h!]
\centering
\includegraphics[width=\linewidth]{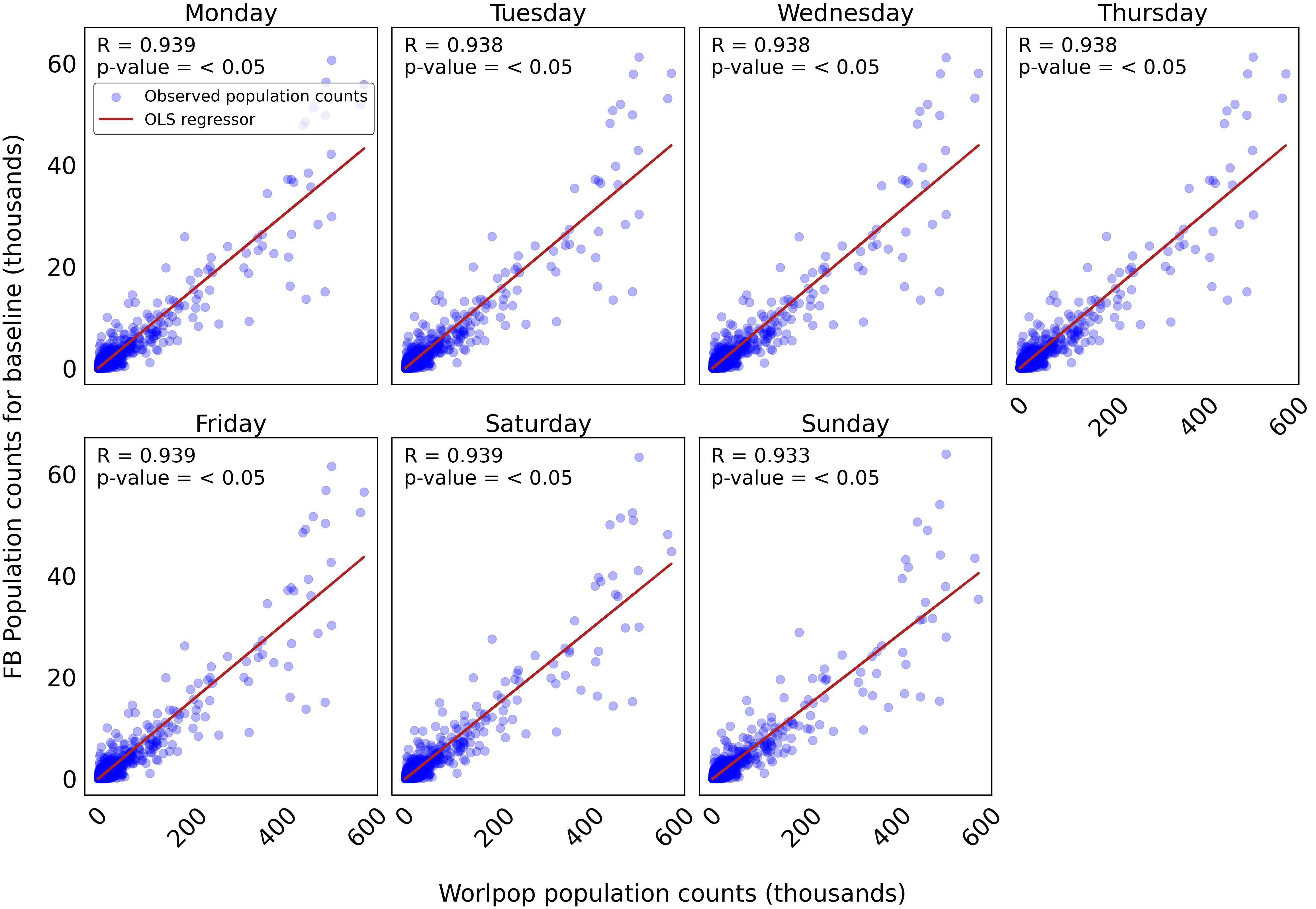}
\caption{Correlation between Worldpop population counts and Facebook Population counts during the baseline period in Colombia, for each day of the week.}
\label{correlation-wp-fbpop-COL}
\end{figure}

\newpage

\subsection{Assessing results of spatial interaction model for imputing missing number of movements in baseline}

\begin{figure}[h!]
\centering
\includegraphics[width=\linewidth]{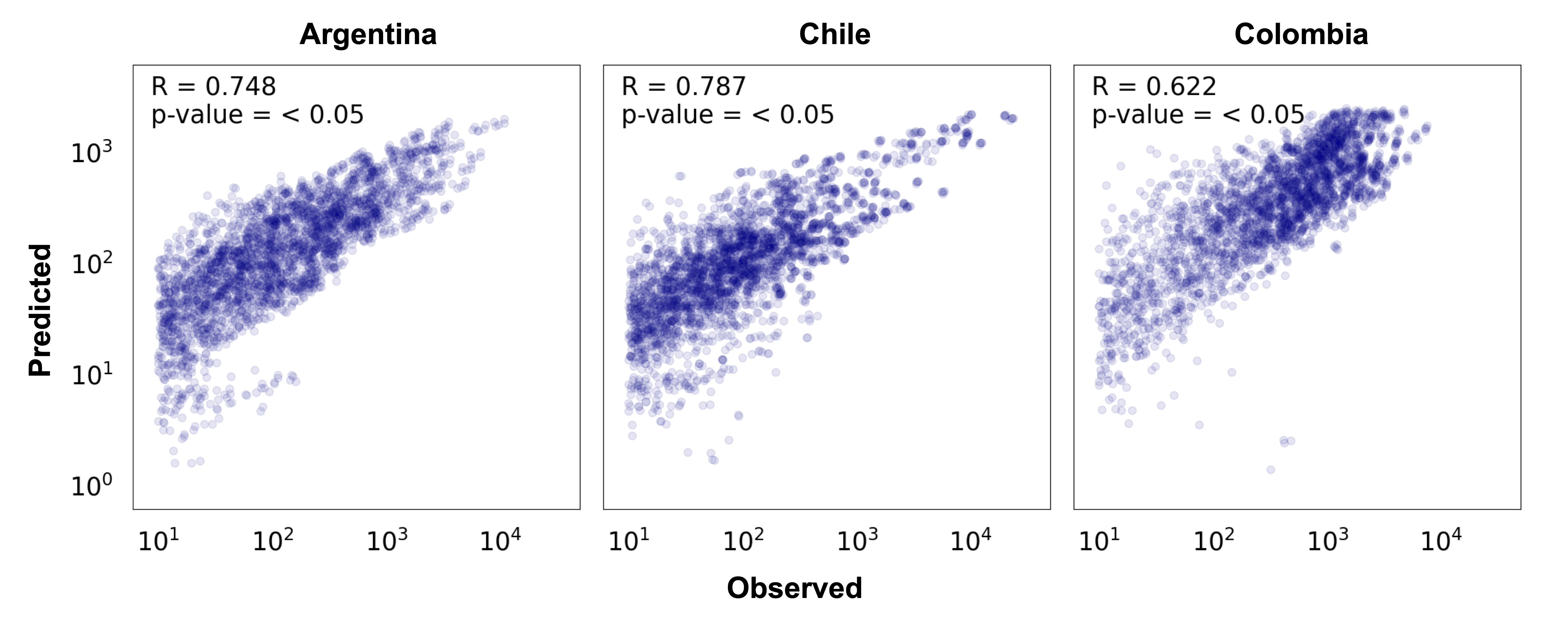}
\caption{Relationship between observed and predicted value of number of movements between origin-destination pairs of Bing tiles during the baseline, for all weekdays.}
\end{figure}

\begin{figure}[h!]
\centering
\includegraphics[width=\linewidth]{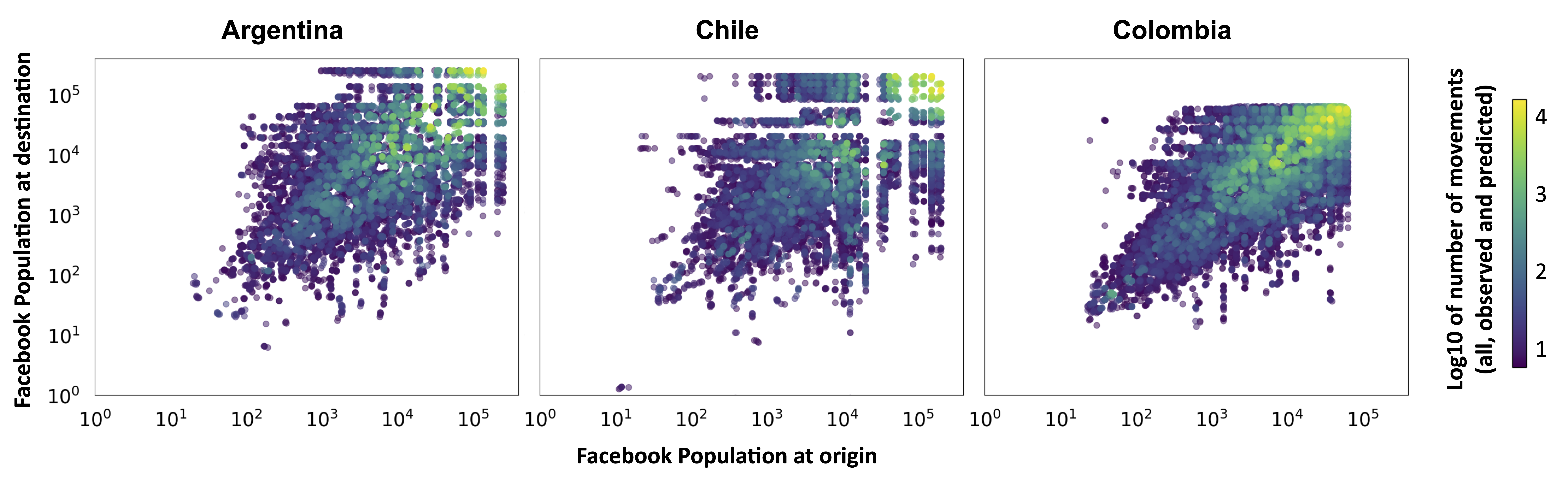}
\caption{Relationship between Facebook Population at the origin and the destination of each population flow during the baseline period. The Facebook Population is considered after the imputation of the Facebook Population baseline data. The colour of the markers represents the volume of the flows during the baseline period, both observed and predicted through the spatial interaction model. Note that the Bing tiles used to record Facebook Movement data in Colombia are smaller than in Argentina and Chile, hence the lower numbers. }
\end{figure}

\newpage

\section{Classification of tiles according to level of urbanisation and socioeconomic deprivation}

\begin{figure}[h!]
\centering
\includegraphics[width=\linewidth]{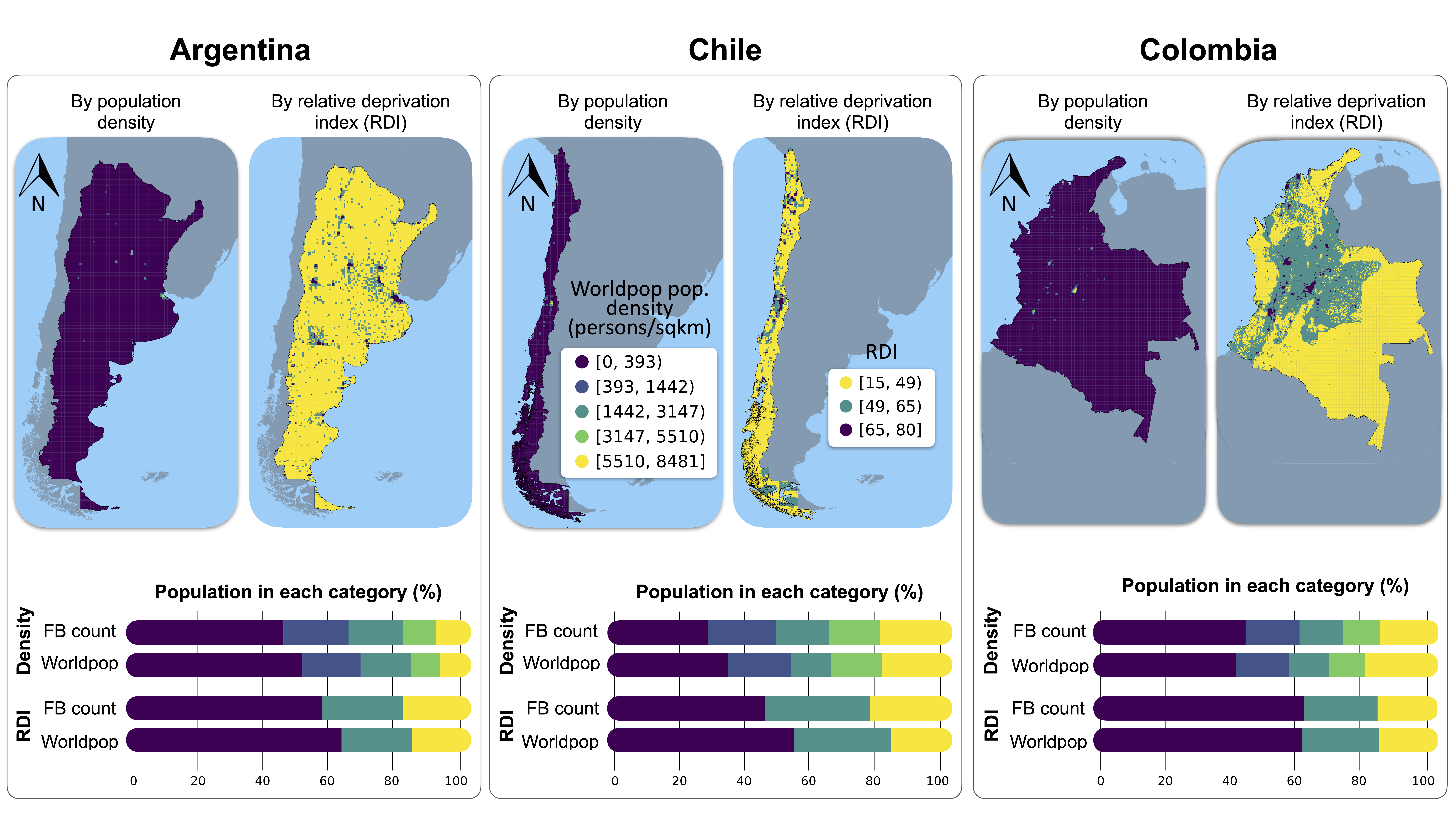}
\caption{Classification of spatial units into categories by population density and by relative deprivation index. Spatial distribution of categories and population share in each category, by country.}
\end{figure}

\begin{figure}[h!]
\centering
\includegraphics[width=\linewidth]{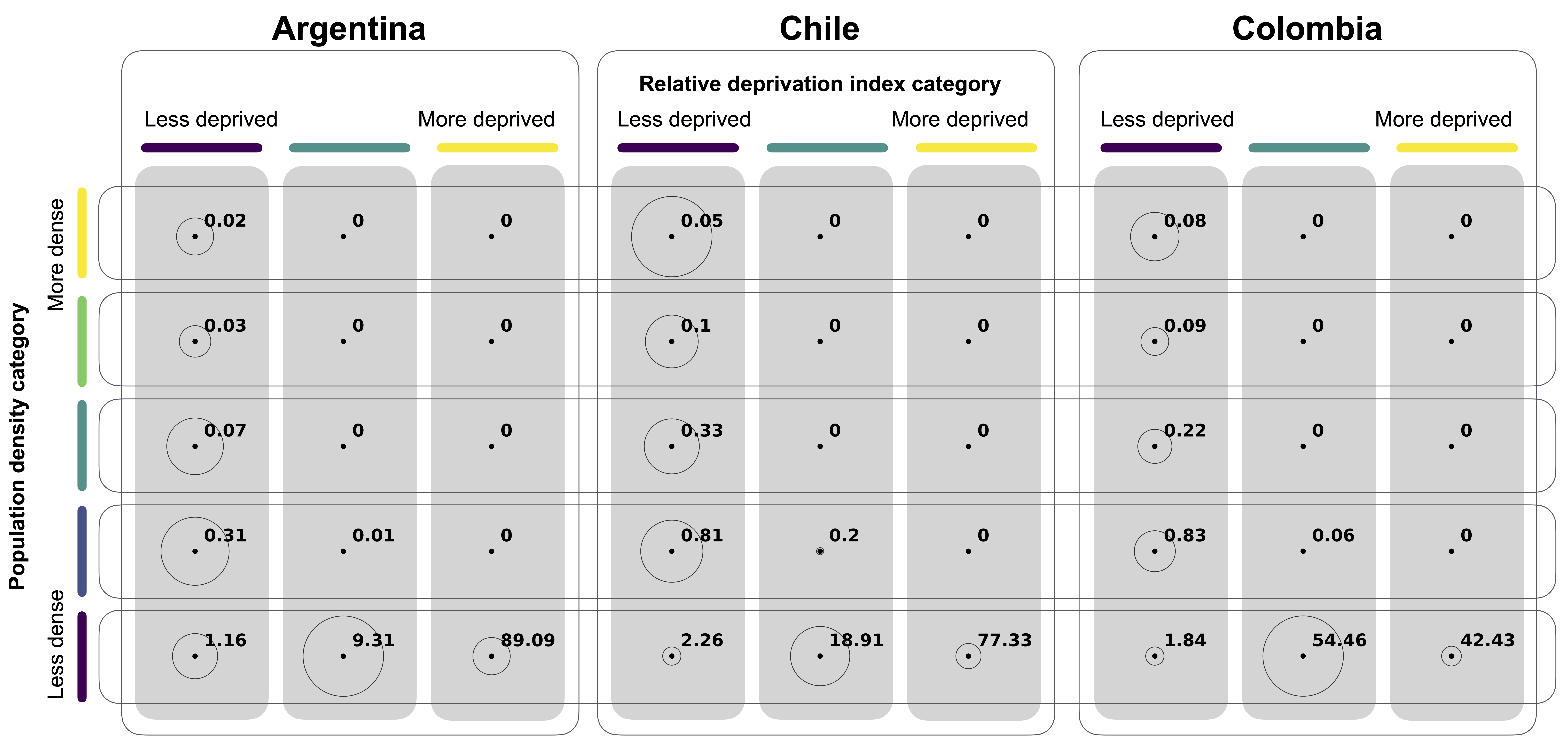}
\caption{Joint distribution of spatial units by population density (rows) and relative deprivation index (RDI) categories (columns), with values representing the percentage of spatial units within each country. Circle sizes indicate the total population corresponding to each category pair.}
\end{figure}

\newpage

\section{Trend analysis}

\subsection{Time series to be model with trend analysis}

\begin{figure}[h!]
\centering
\includegraphics[width=\linewidth]{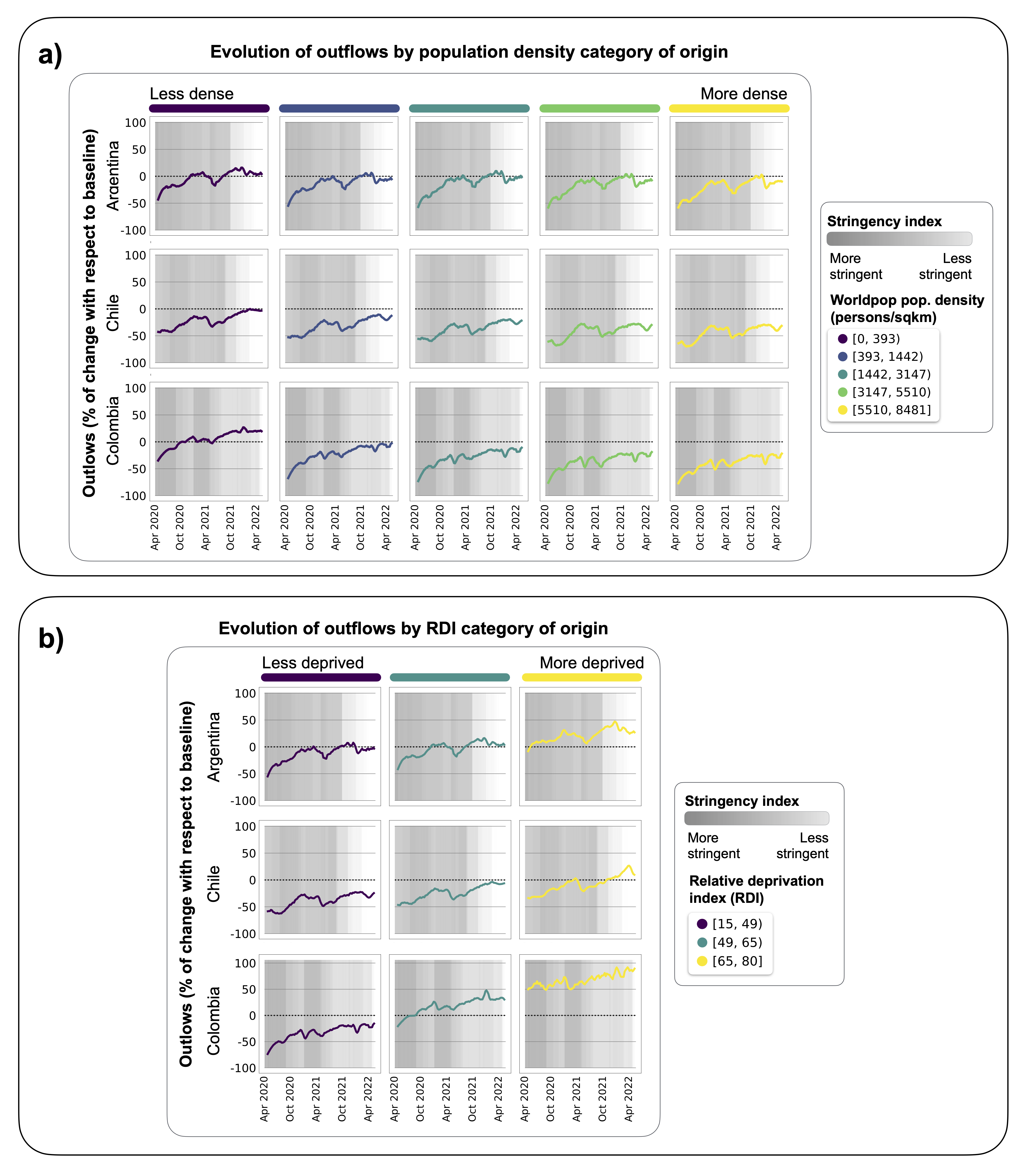}
\caption{Time evolution of percentage change of outflows relative to pre-pandemic baseline levels, by country and by (a) population density category and (b) relative deprivation category.}
\label{fig-classification}
\end{figure}

\subsection{Mixed effects model specification}

\begin{table}[h!]
\centering
\label{tab:model-specs}
\begin{tabular}{cccccccc}
\toprule
\textbf{Model} & 
\makecell{\textbf{Time} \\ \textbf{(Linear)}} & 
\makecell{\textbf{Time}$^2$} & 
\makecell{\textbf{Origin category} \\ \textbf{(Fixed)}} & 
\makecell{\textbf{Stringency} \\ \textbf{index}} & 
\makecell{\textbf{Random} \\ \textbf{intercept}} & 
\makecell{\textbf{Random slope} \\ \textbf{(Time)}} & 
\makecell{\textbf{Random slope} \\ \textbf{(Time}$^2$\textbf{)}} \\
\midrule
1 & \checkmark & & & (\checkmark) & & & \\
2 & \checkmark & & \checkmark & (\checkmark) & & & \\
3 & \checkmark & & & (\checkmark) & \checkmark & & \\
4 & \checkmark & & & (\checkmark) & & \checkmark & \\
5 & \checkmark & & & (\checkmark) & \checkmark & \checkmark & \\
6 & \checkmark & \checkmark & & (\checkmark) & \checkmark & \checkmark & \\
7 & \checkmark & \checkmark & & (\checkmark) & \checkmark & \checkmark & \checkmark \\
\end{tabular}
\caption{Summary of mixed-effects model specifications. There is a total of 14. Each row in the table represents a pair of models: one that includes the stringency index as a covariate and one that excludes it. This is denoted by a bracketed check mark [(\checkmark)]. All other check marks [\checkmark] indicate terms that are included in both models of the pair. Random effects are considered according to the population density or the Relative Deprivation Index (RDI) of the origin location of outflows.}
\end{table}

\newpage

\subsection{Relative change in outflows}

\begin{table}[h!]
\centering
\includegraphics[width=\linewidth]{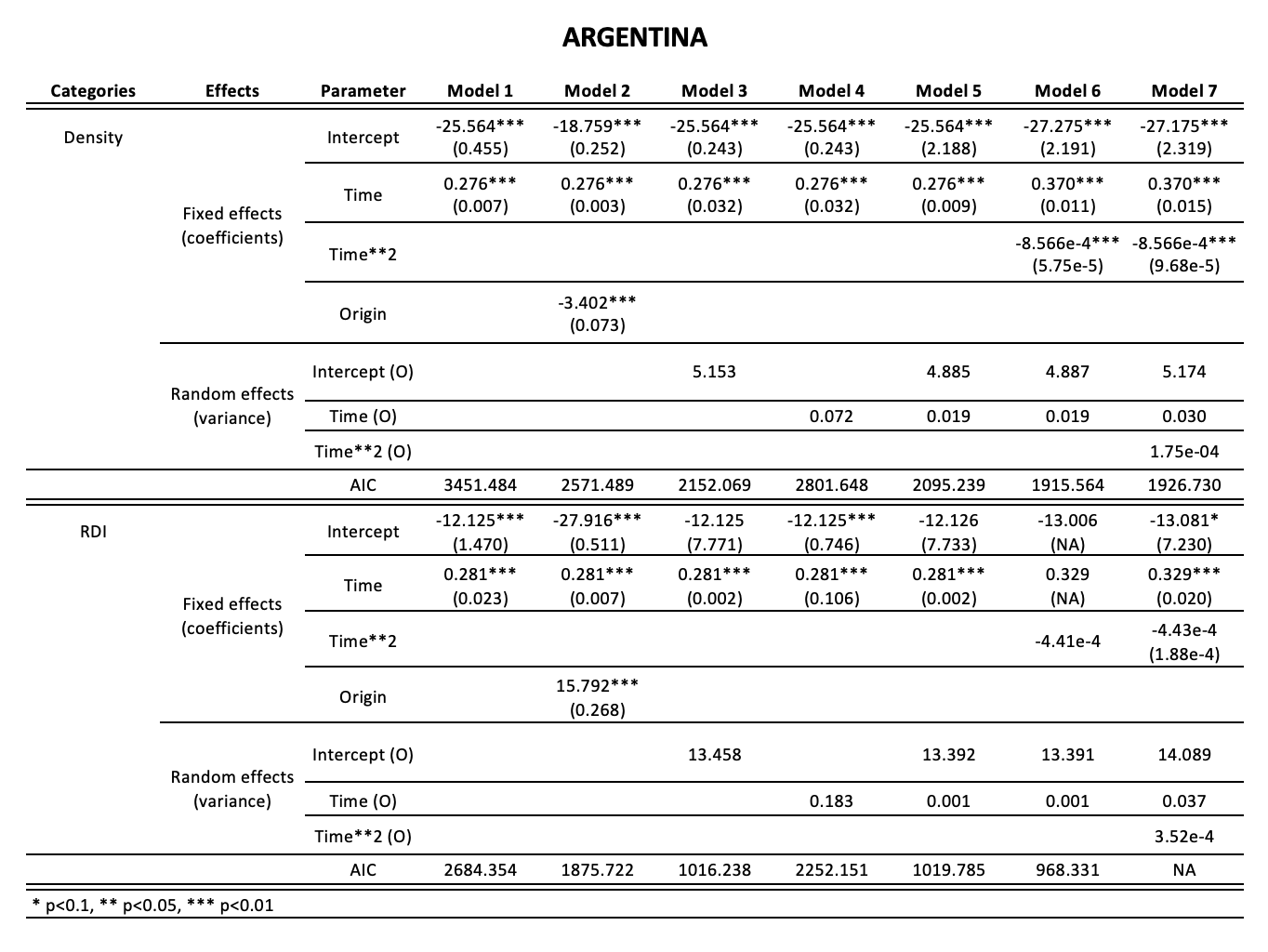}
\caption{Results of mixed-effects models for trend analysis of relative change in outflows in Argentina. Random effects according to the population density or the Relative Deprivation Index (RDI) of the origin location of outflows.}
\end{table}

\newpage

\begin{table}[h!]
\centering
\includegraphics[width=\linewidth]{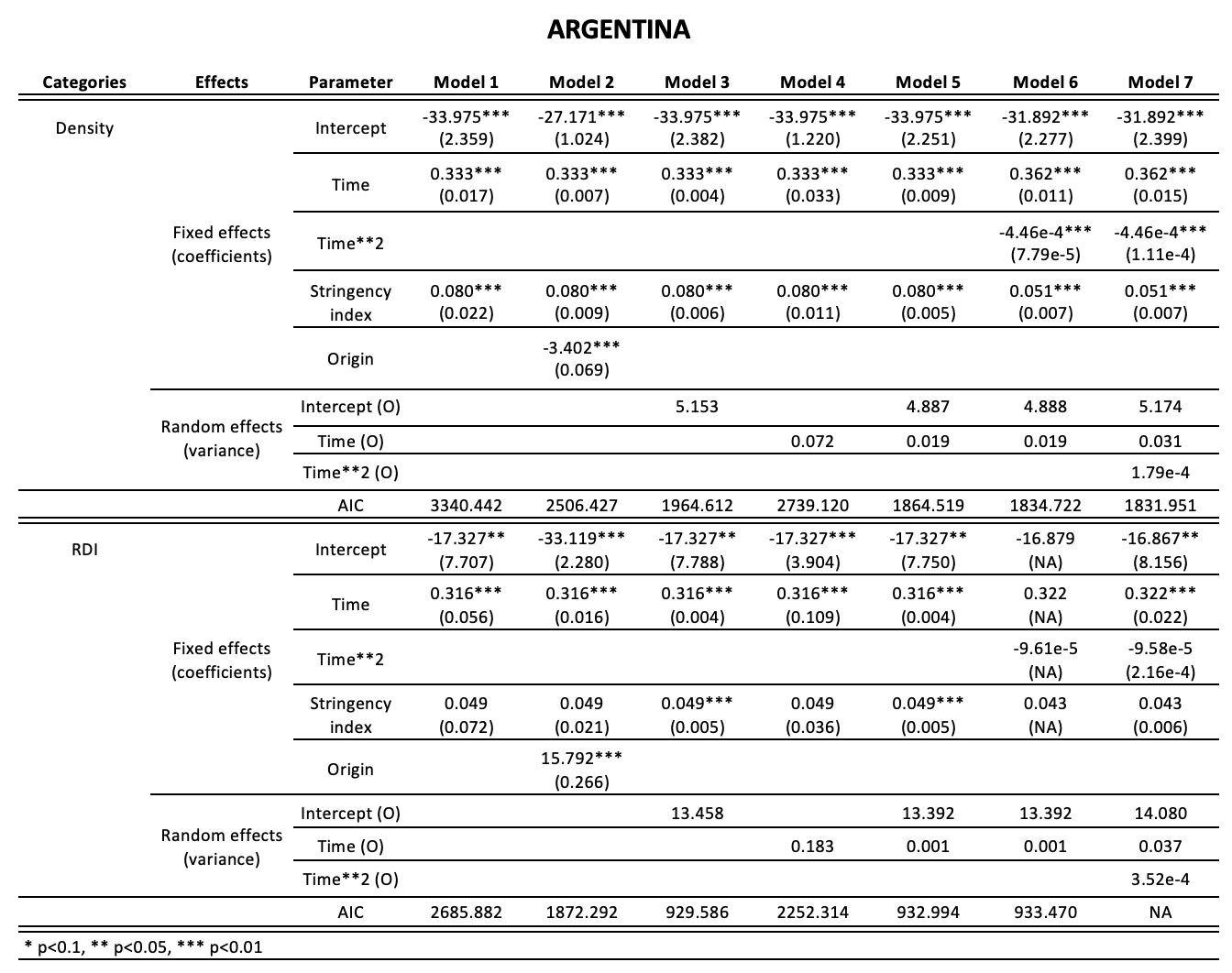}
\caption{Results of mixed-effects models for trend analysis of relative change in outflows in Argentina including stringency index as a predictor variable. Random effects according to the population density or the Relative Deprivation Index (RDI) of the origin location of outflows.}
\label{tab-modelsARG-stringency}
\end{table}

\newpage

\begin{table}[h!]
\centering
\includegraphics[width=\linewidth]{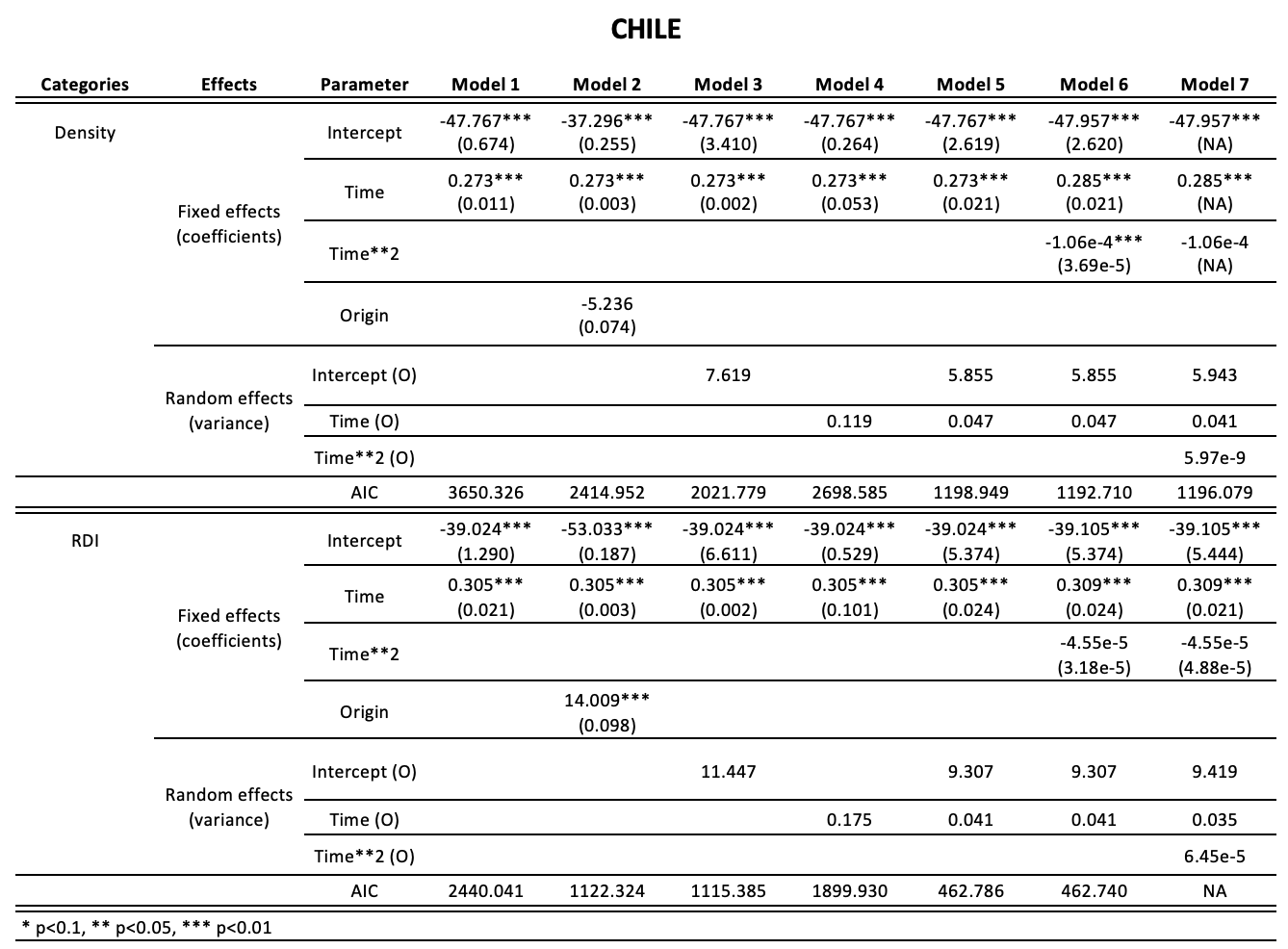}
\caption{Results of mixed-effects models for trend analysis of relative change in outflows in Chile. Random effects according to the population density or the Relative Deprivation Index (RDI) of the origin location of outflows.}
\label{tab-modelsCHL}
\end{table}

\newpage

\begin{table}[h!]
\centering
\includegraphics[width=\linewidth]{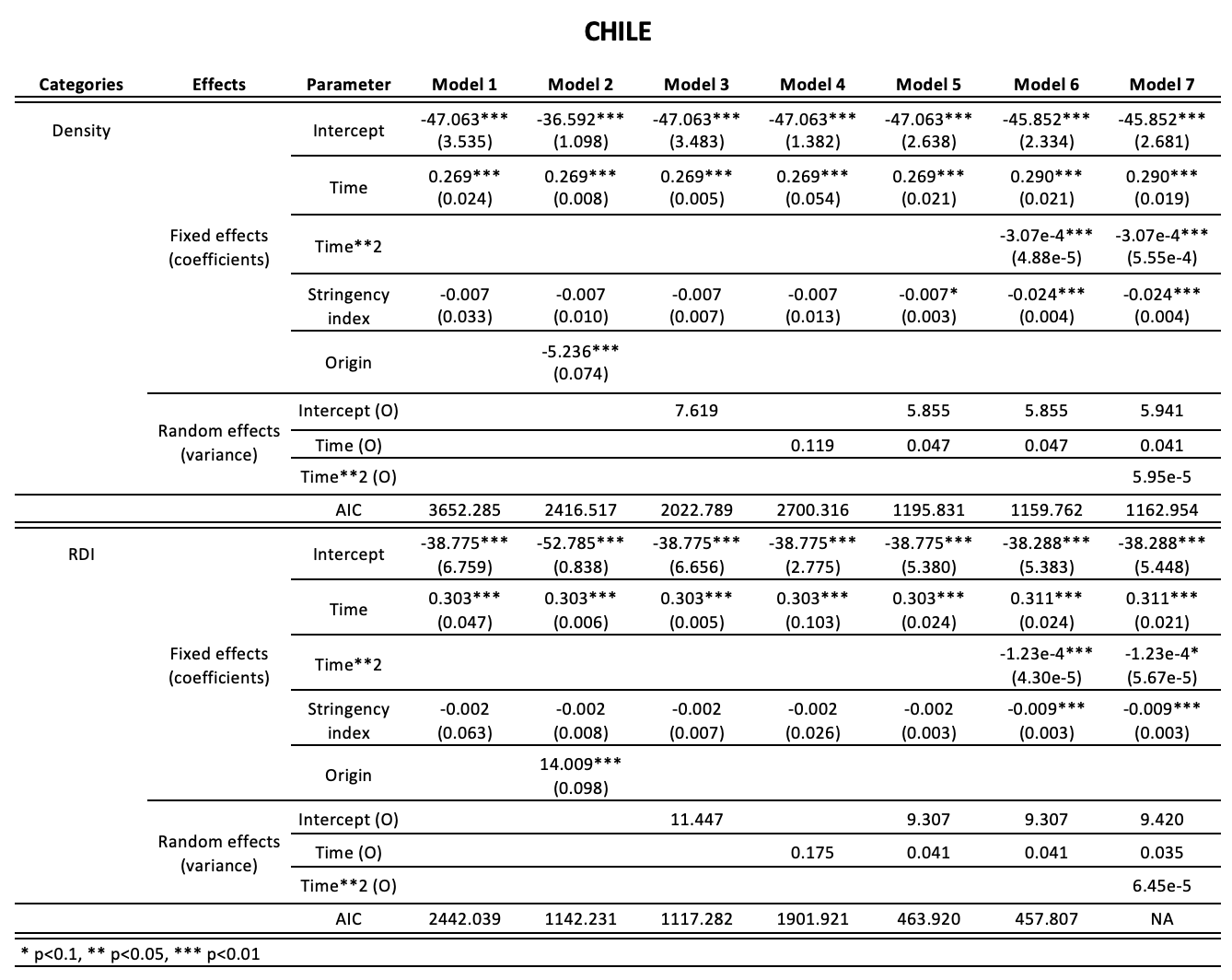}
\caption{Results of mixed-effects models for trend analysis of relative change in outflows in Chile including stringency index as a predictor variable. Random effects according to the population density or the Relative Deprivation Index (RDI) of the origin location of outflows.}
\label{tab-modelsCHL-stringency}
\end{table}

\newpage

\begin{table}[h!]
\centering
\includegraphics[width=\linewidth]{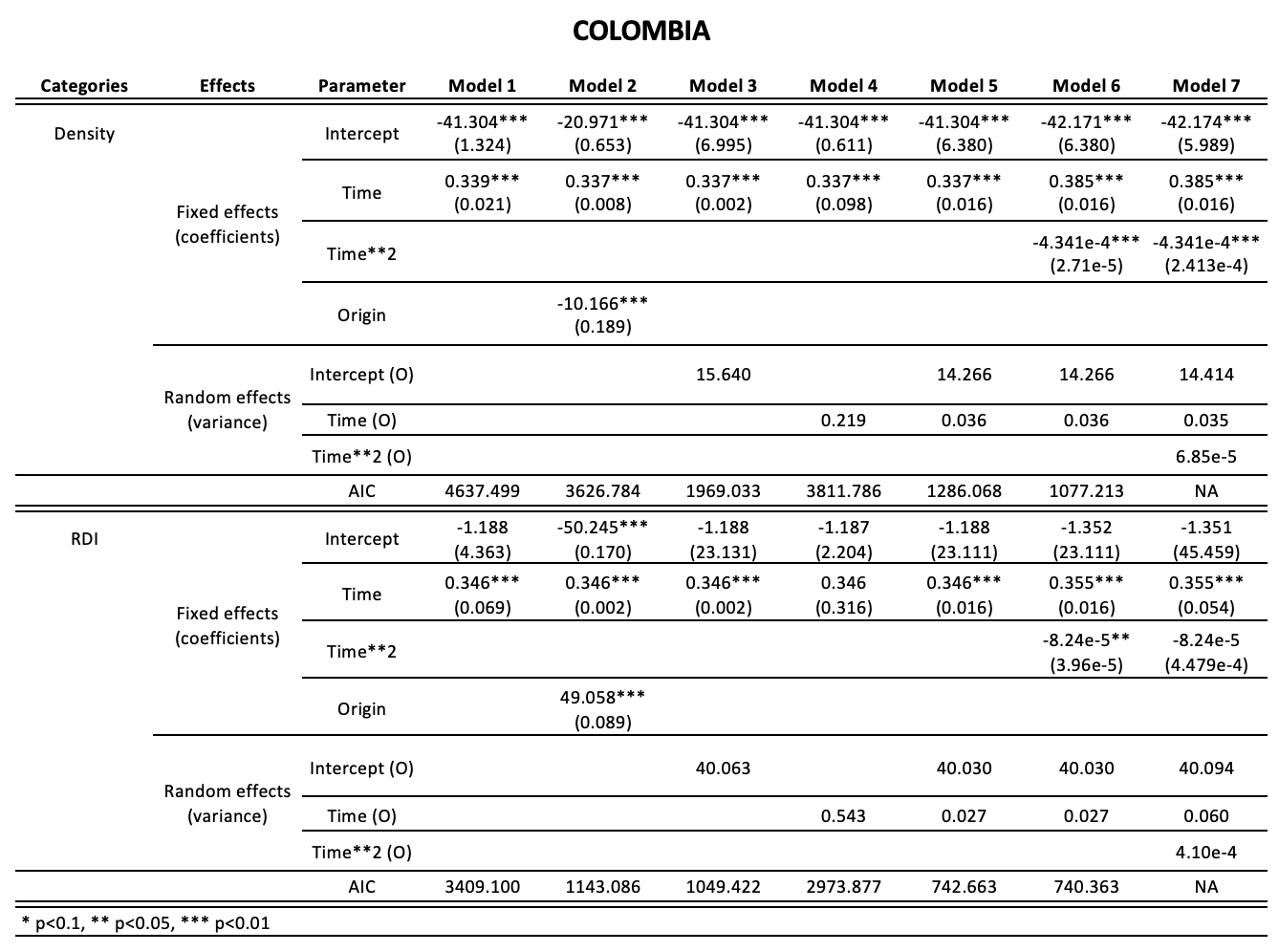}
\caption{Results of mixed-effects models for trend analysis of relative change in outflows in Colombia. Random effects according to the population density or the Relative Deprivation Index (RDI) of the origin location of outflows.}
\label{tab-modelsCOL}
\end{table}

\newpage

\begin{table}[h!]
\centering
\includegraphics[width=\linewidth]{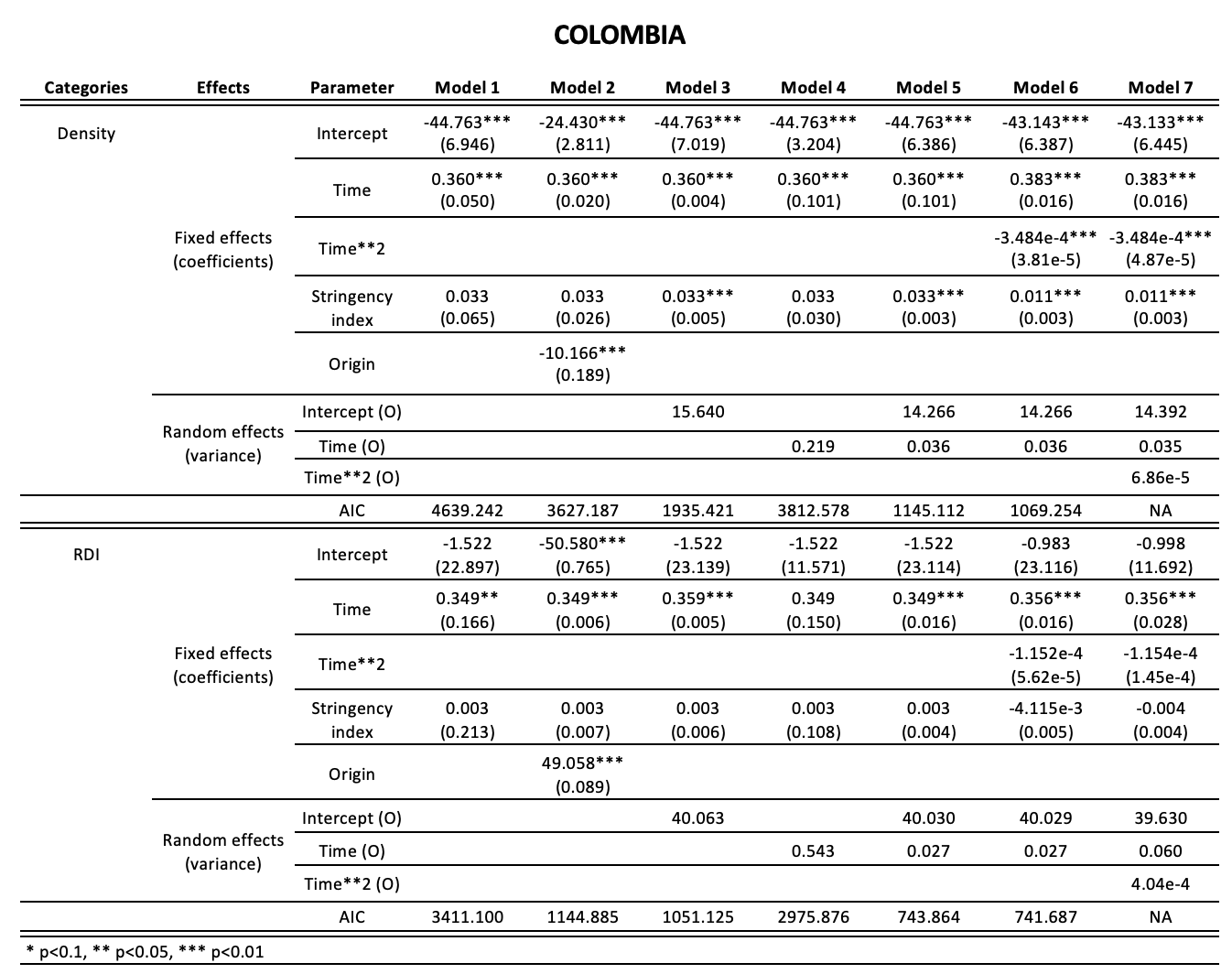}
\caption{Results of mixed-effects models for trend analysis of relative change in outflows in Colombia including stringency index as a predictor variable. Random effects according to the population density or the Relative Deprivation Index (RDI) of the origin location of outflows.}
\label{tab-modelsCOL-stringency}
\end{table}

\newpage

\subsection{Relative change in netflows involving capital}

\begin{table}[h!]
\centering
\includegraphics[width=\linewidth]{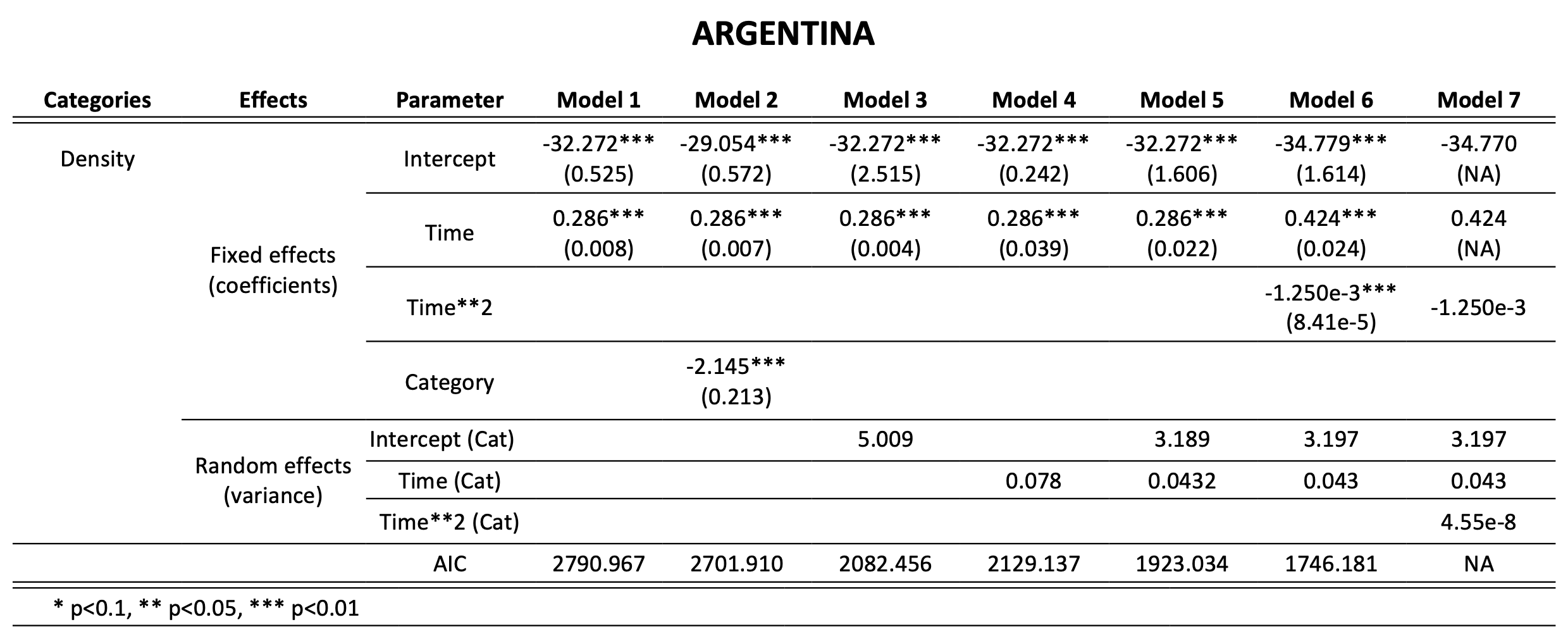}
\caption{Results of mixed-effects models for trend analysis of relative change in netflows involving urban region surrounding capital in Argentina. Netflows are defined as inflows to an urban core minus outflows from the urban core. Random effects according to the population density or the Relative Deprivation Index (RDI) of the origin location of outflows.}
\label{tab-modelsARG-capital}
\end{table}

\begin{table}[h!]
\centering
\includegraphics[width=\linewidth]{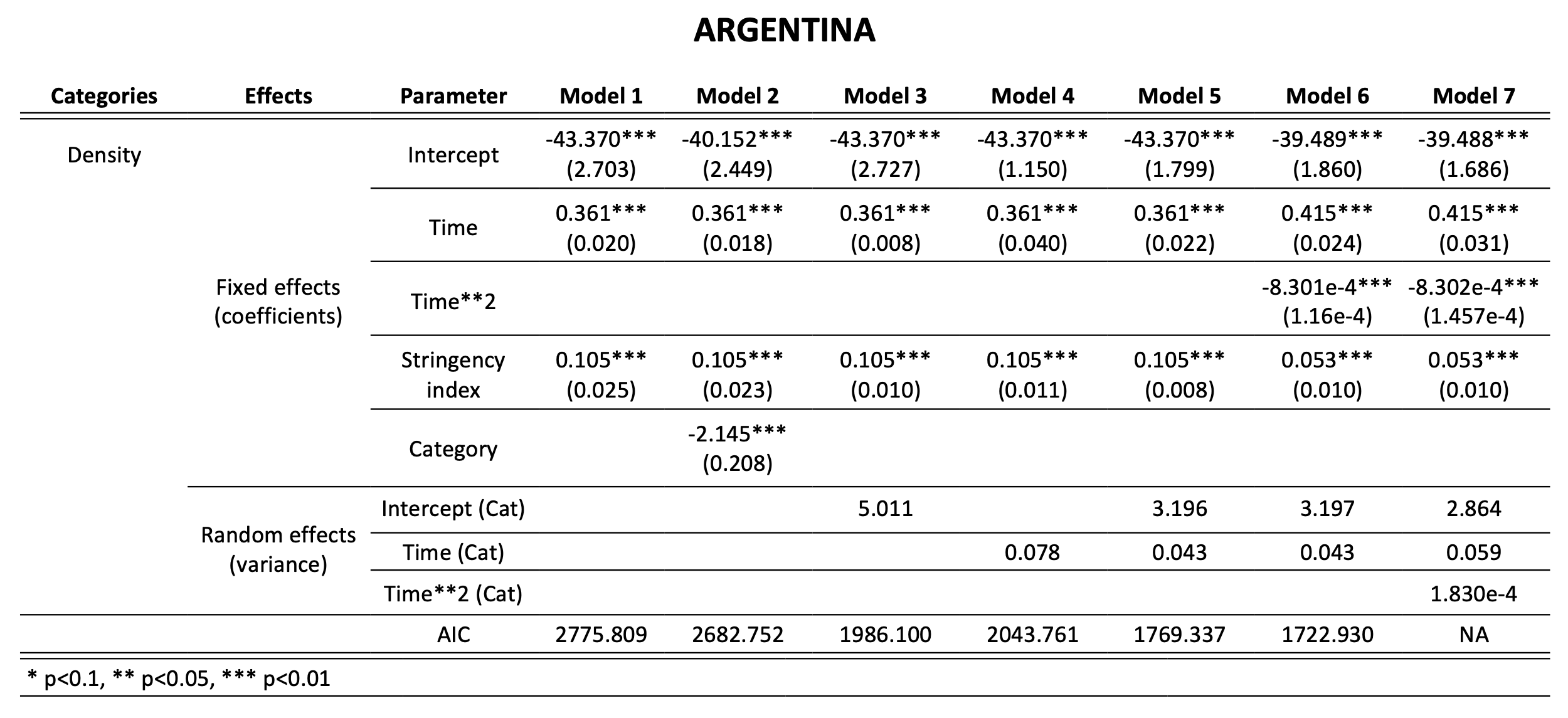}
\caption{Results of mixed-effects models for trend analysis of relative change in netflows involving urban region surrounding capital in Argentina including stringency index as a predictor variable. Netflows are defined as inflows to an urban core minus outflows from the urban core. Random effects according to the population density or the Relative Deprivation Index (RDI) of the origin location of outflows.}
\label{tab-modelsARG-capital-stringency}
\end{table}

\newpage

\begin{table}[h!]
\centering
\includegraphics[width=\linewidth]{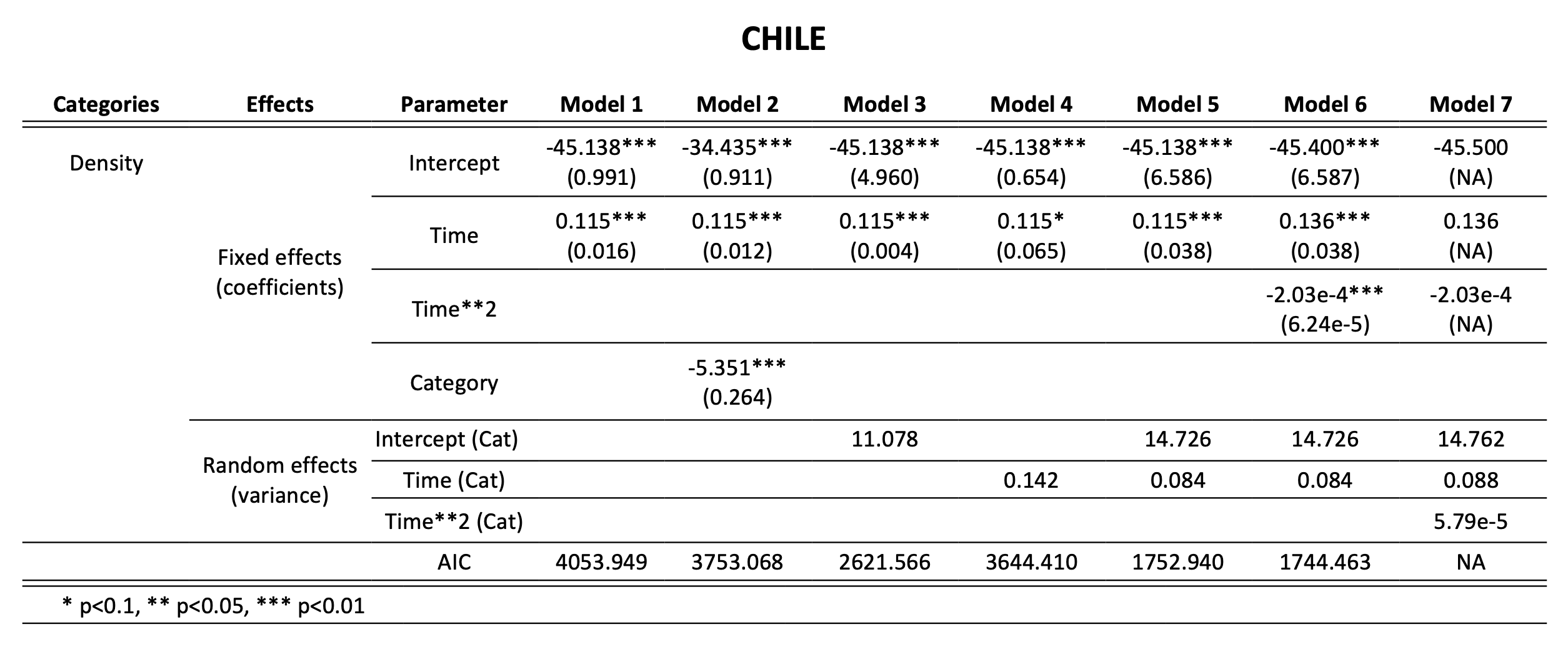}
\caption{Results of mixed-effects models for trend analysis of relative change in netflows involving urban region surrounding capital in Chile. Netflows are defined as inflows to an urban core minus outflows from the urban core. Random effects according to the population density or the Relative Deprivation Index (RDI) of the origin location of outflows.}
\label{tab-modelsCHL-capital}
\end{table}

\begin{table}[h!]
\centering
\includegraphics[width=\linewidth]{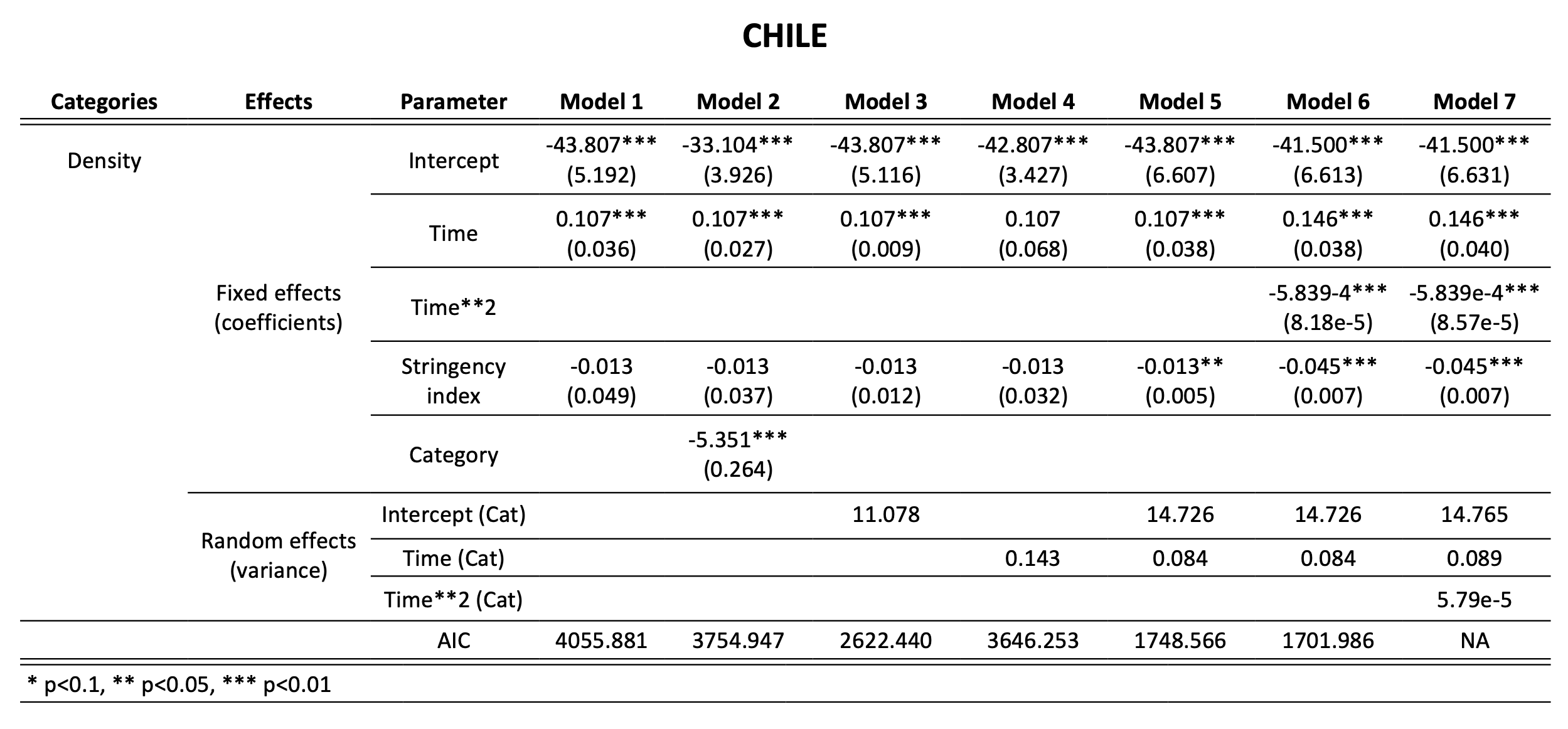}
\caption{Results of mixed-effects models for trend analysis of relative change in netflows involving urban region surrounding capital in Chile including stringency index as a predictor variable. Netflows are defined as inflows to an urban core minus outflows from the urban core. Random effects according to the population density or the Relative Deprivation Index (RDI) of the origin location of outflows.}
\label{tab-modelsCHL-capital-stringency}
\end{table}

\newpage

\begin{table}[h!]
\centering
\includegraphics[width=\linewidth]{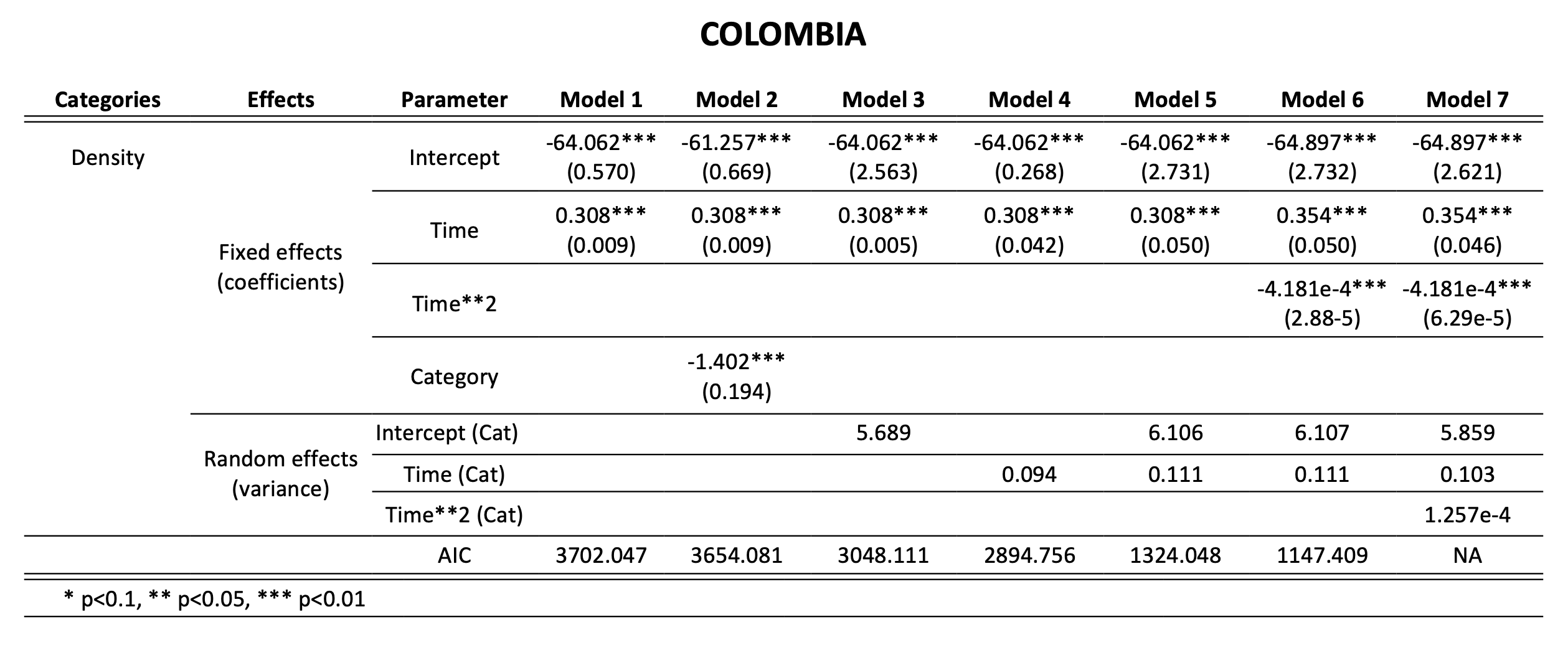}
\caption{Results of mixed-effects models for trend analysis of relative change in netflows involving urban region surrounding capital in Colombia. Netflows are defined as inflows to an urban core minus outflows from the urban core. Random effects according to the population density or the Relative Deprivation Index (RDI) of the origin location of outflows.}
\label{tab-modelsCOL-capital}
\end{table}

\begin{table}[h!]
\centering
\includegraphics[width=\linewidth]{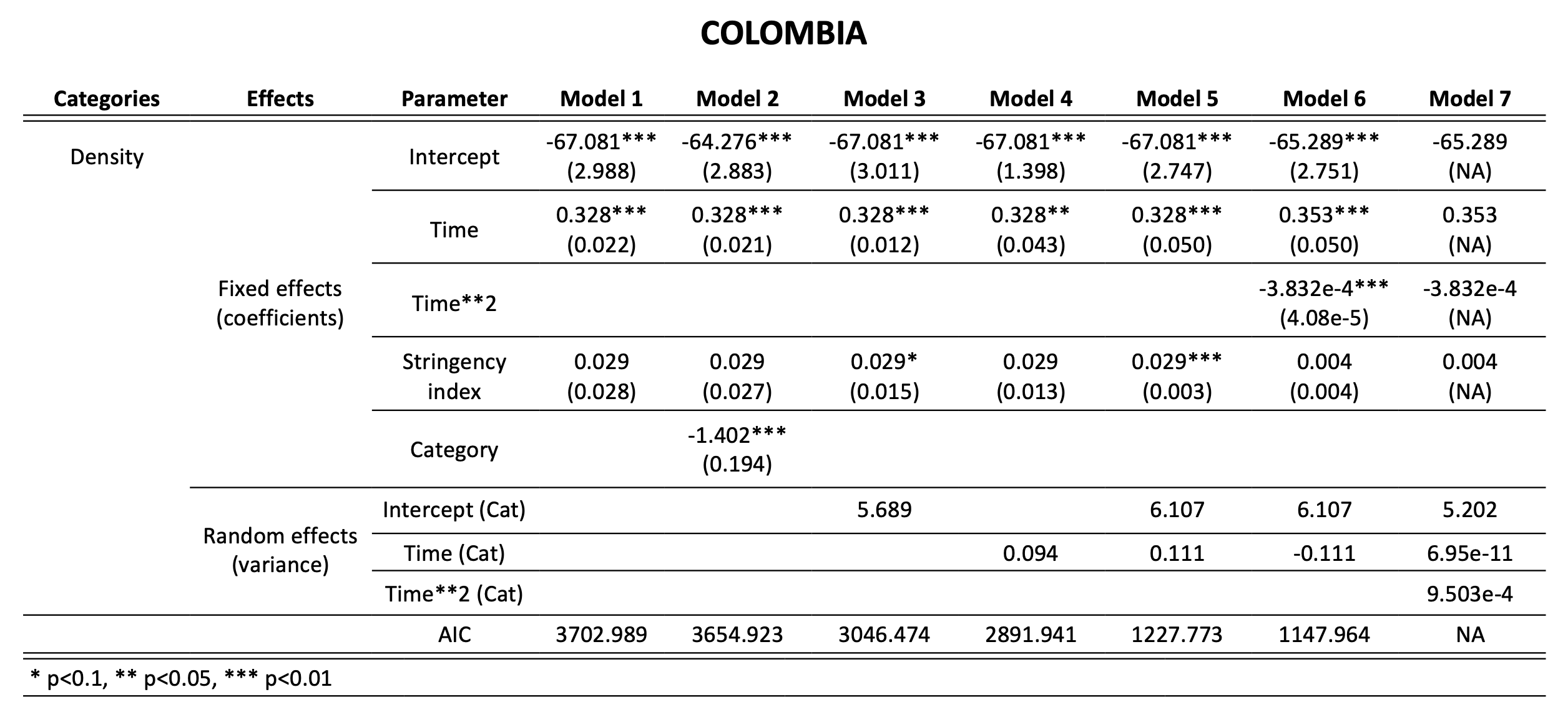}
\caption{Results of mixed-effects models for trend analysis of relative change in netflows involving urban region surrounding capital in Colombia including stringency index as a predictor variable. Netflows are defined as inflows to an urban core minus outflows from the urban core. Random effects according to the population density or the Relative Deprivation Index (RDI) of the origin location of outflows.}
\label{tab-modelsCOL-capital-stringency}
\end{table}

\newpage

\subsection{Relative change in netflows involving all urban regions}

\begin{table}[h!]
\centering
\includegraphics[width=\linewidth]{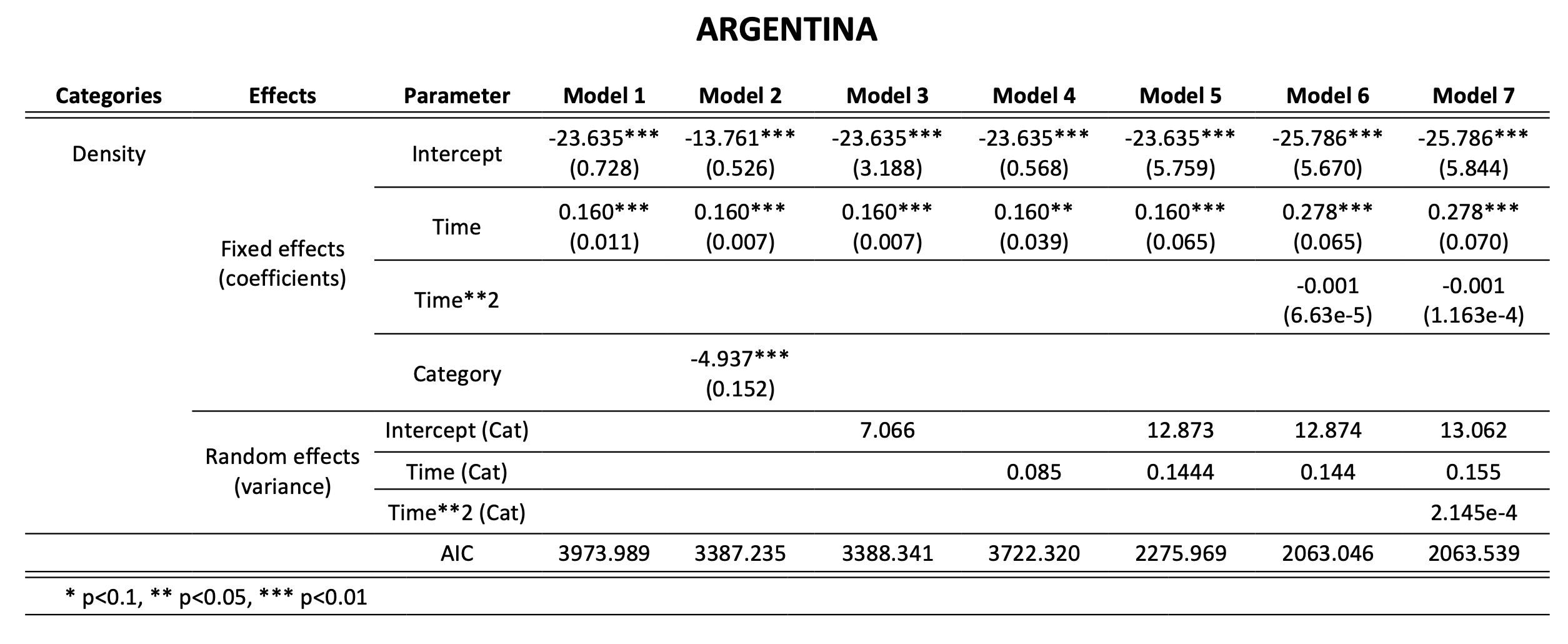}
\caption{Results of mixed-effects models for trend analysis of relative change in netflows involving all urban regions in Argentina. Netflows are defined as inflows to an urban core minus outflows from the urban core. Random effects according to the population density or the Relative Deprivation Index (RDI) of the origin location of outflows.}
\label{tab-modelsARG-fuas}
\end{table}

\begin{table}[h!]
\centering
\includegraphics[width=\linewidth]{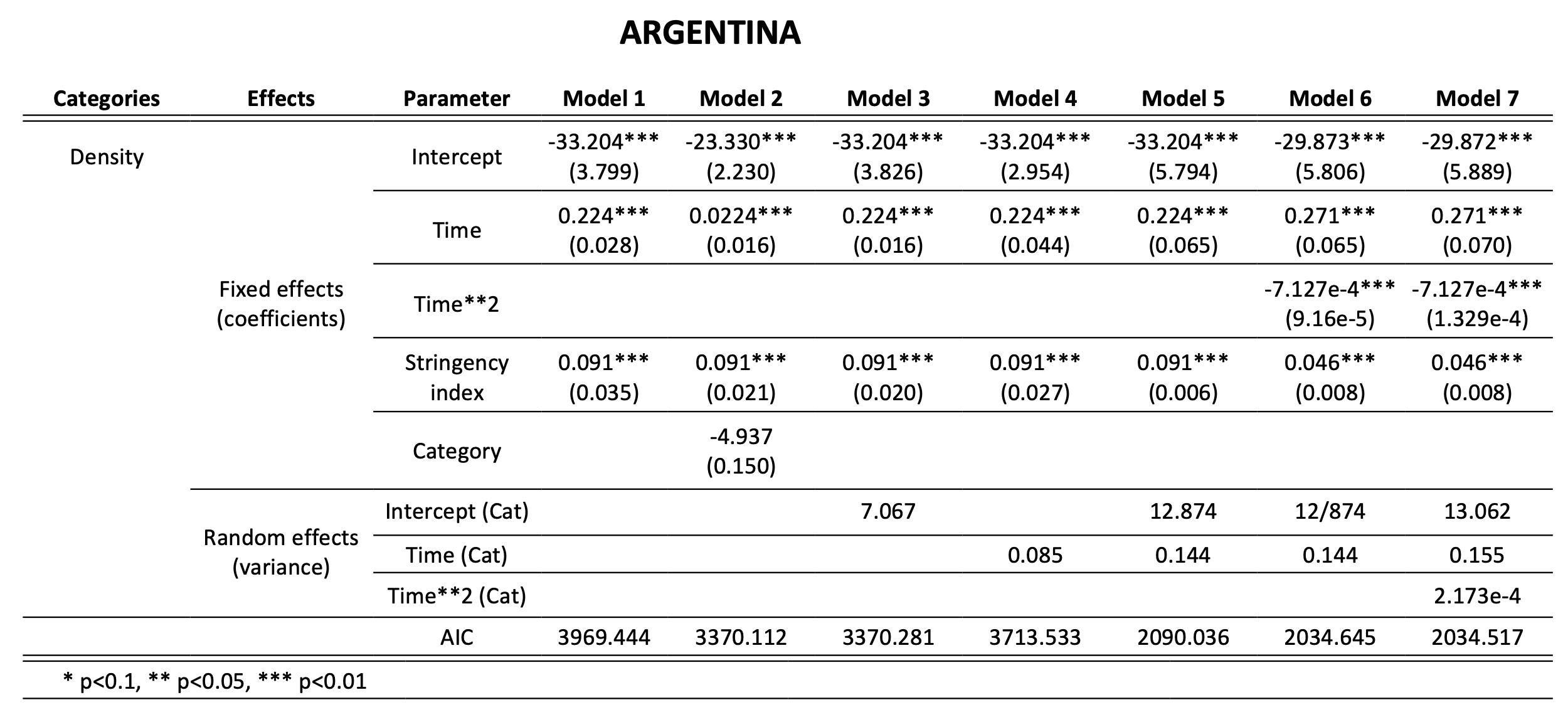}
\caption{Results of mixed-effects models for trend analysis of relative change in netflows involving all urban regions in Argentina including stringency index as a predictor variable. Netflows are defined as inflows to an urban core minus outflows from the urban core. Random effects according to the population density or the Relative Deprivation Index (RDI) of the origin location of outflows.}
\label{tab-modelsARG-fuas-stringency}
\end{table}

\newpage

\begin{table}[h!]
\centering
\includegraphics[width=\linewidth]{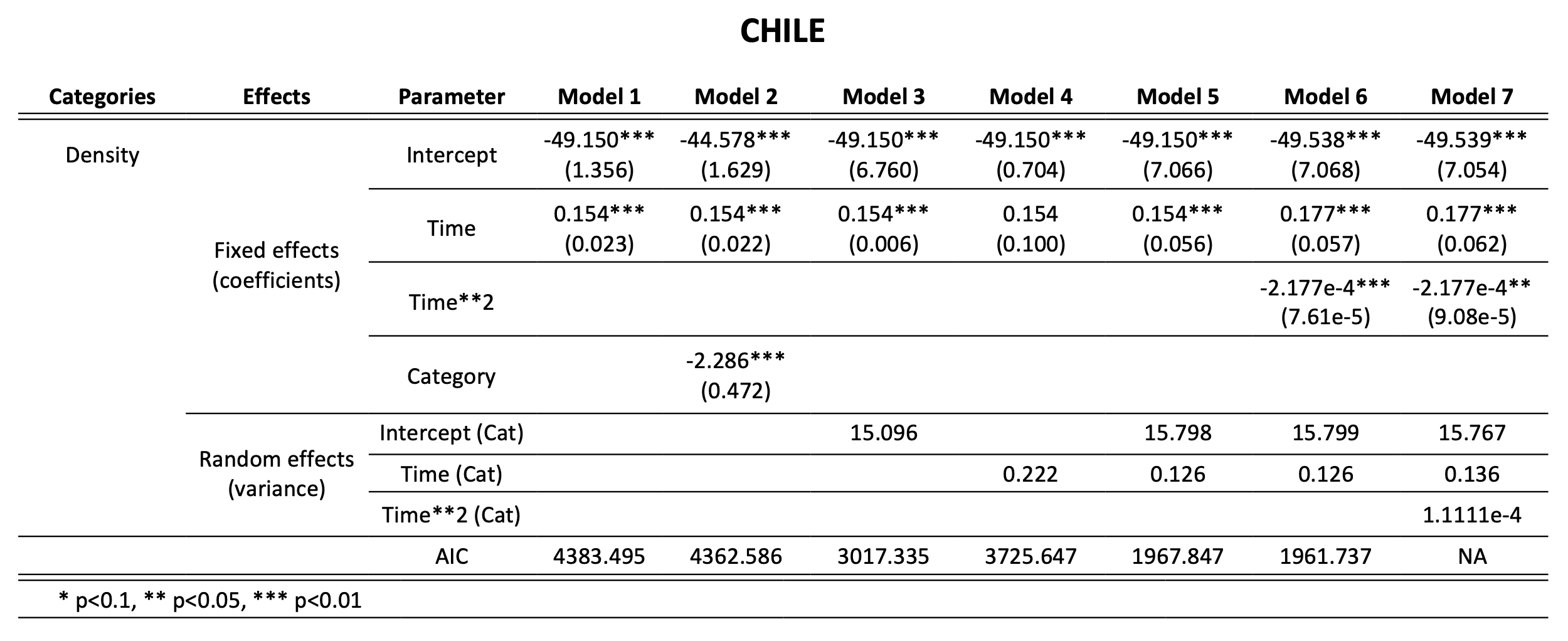}
\caption{Results of mixed-effects models for trend analysis of relative change in netflows involving all urban regions in Chile. Netflows are defined as inflows to an urban core minus outflows from the urban core. Random effects according to the population density or the Relative Deprivation Index (RDI) of the origin location of outflows.}
\label{tab-modelsCHL-fuas}
\end{table}

\begin{table}[h!]
\centering
\includegraphics[width=\linewidth]{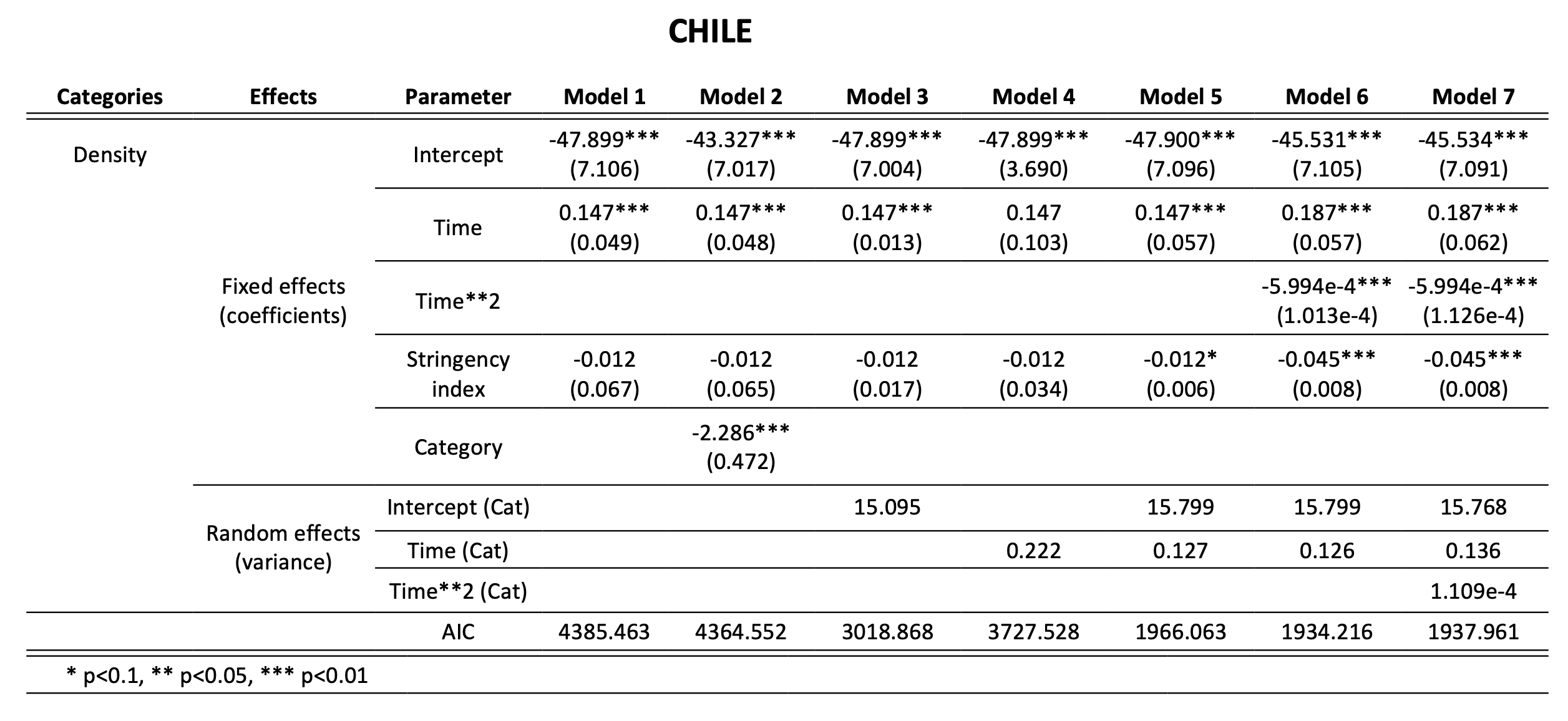}
\caption{Results of mixed-effects models for trend analysis of relative change in netflows involving all urban regions in Chile including stringency index as a predictor variable. Netflows are defined as inflows to an urban core minus outflows from the urban core. Random effects according to the population density or the Relative Deprivation Index (RDI) of the origin location of outflows.}
\label{tab-modelsCHL-fuas-stringency}
\end{table}

\newpage

\begin{table}[h!]
\centering
\includegraphics[width=\linewidth]{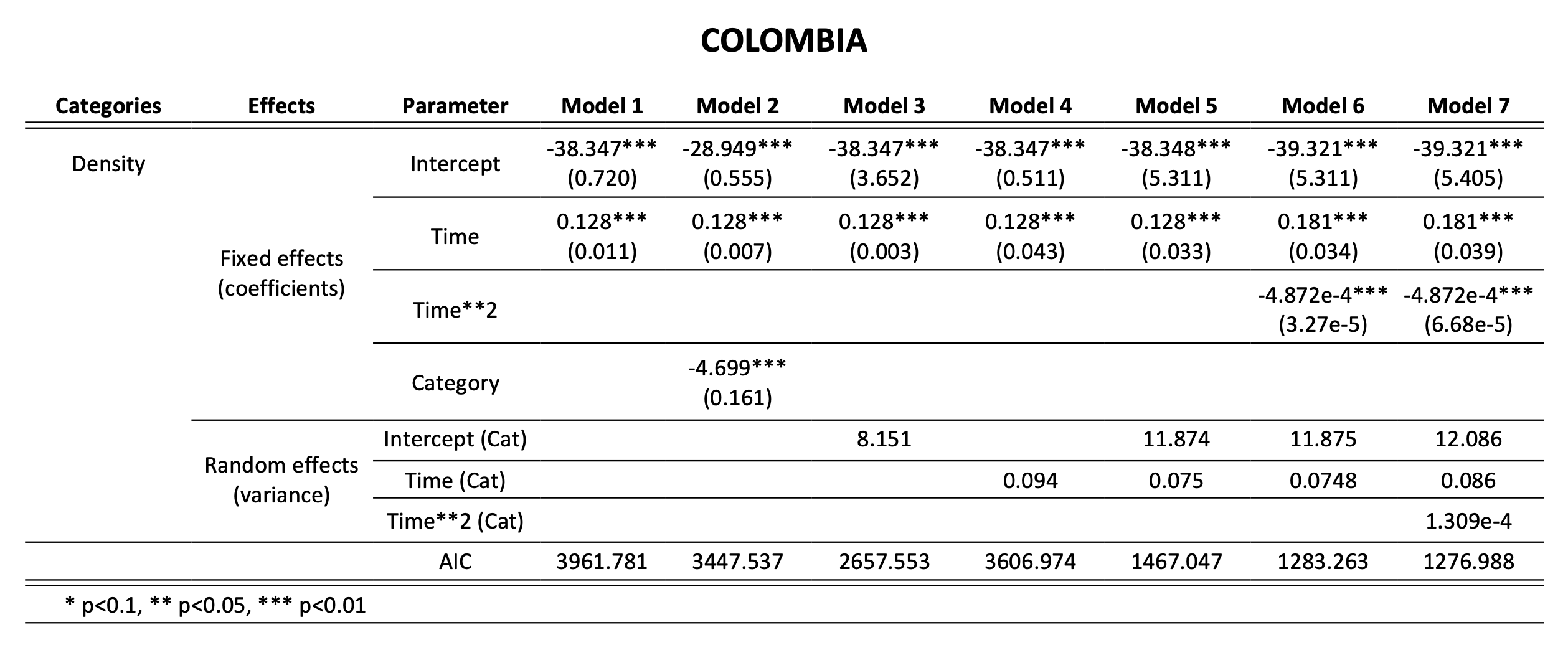}
\caption{Results of mixed-effects models for trend analysis of relative change in netflows involving all urban regions in Colombia. Netflows are defined as inflows to an urban core minus outflows from the urban core. Random effects according to the population density or the Relative Deprivation Index (RDI) of the origin location of outflows.}
\label{tab-modelsCOL-fuas}
\end{table}

\begin{table}[h!]
\centering
\includegraphics[width=\linewidth]{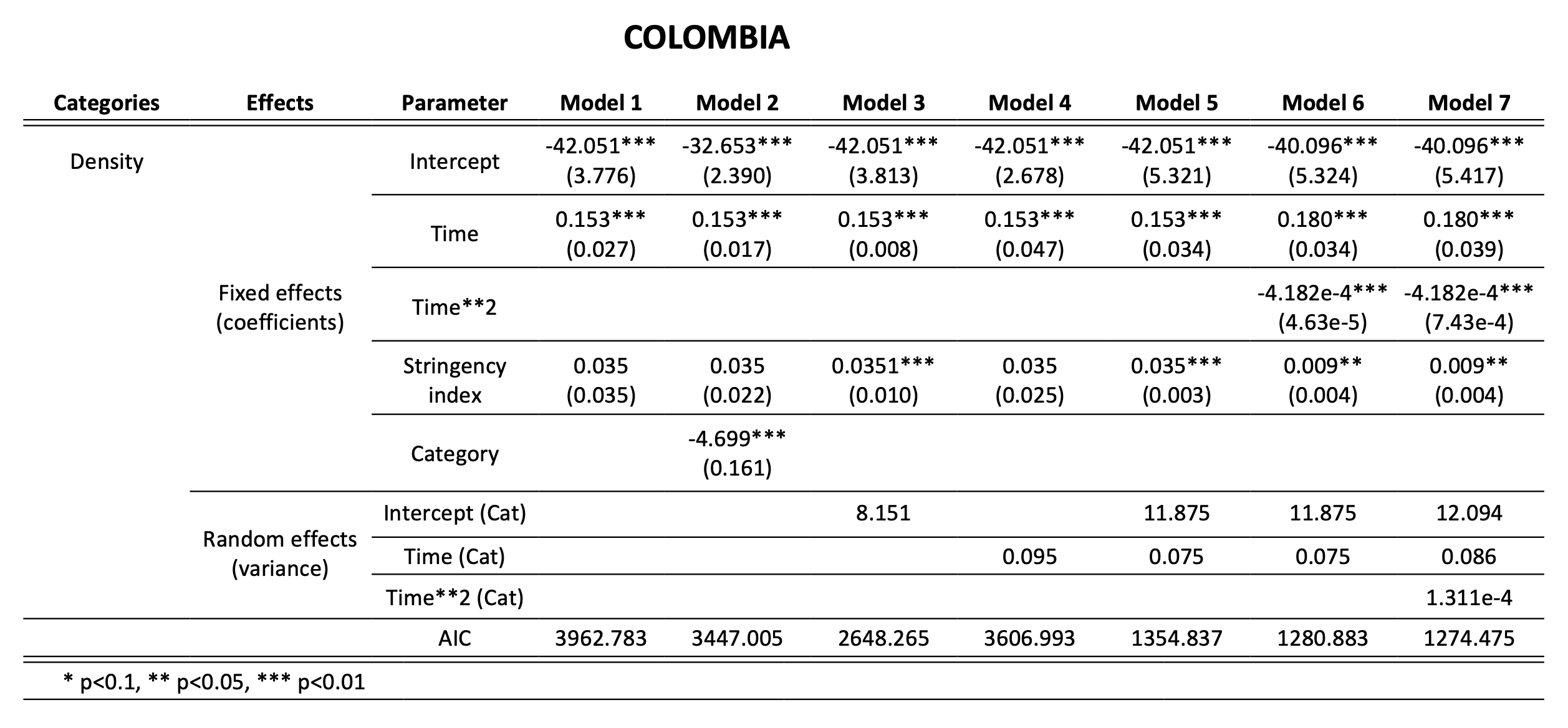}
\caption{Results of mixed-effects models for trend analysis of relative change in netflows involving all urban regions in Colombia including stringency index as a predictor variable. Netflows are defined as inflows to an urban core minus outflows from the urban core. Random effects according to the population density or the Relative Deprivation Index (RDI) of the origin location of outflows.}
\label{tab-modelsCOL-fuas-stringency}
\end{table}

\newpage

\section{Recovery times}

\subsection{Deriving recovery time formula (equation 7)}

We begin by assuming a quadratic model to describe the evolution of mobility levels over time relative to the baseline:
\begin{equation}
    y(t) = a + b\cdot t + c\cdot t^2
\end{equation}\label{model-trend}

The coefficients \( a \), \( b \), and \( c \) can be expressed in terms of the parameters from equation (6) in the main text. For instance, under Model 6, we have \( a = \beta_0 + b_0 \), \( b = \beta_1 + b_{1i} \), and \( c = \beta_2 \).

The recovery time to a threshold \( \alpha \), denoted by \( t_{\alpha} \), is defined as the time it takes for \( y(t) \) to recover by \( \alpha\% \). If the mobility level relative to the baseline at time \( t = 0 \) is \( a \), then \( t_{\alpha} \) is the time at which
\[
y(t_{\alpha}) = \left(1 - \frac{\alpha}{100}\right) a.
\]

Substituting this into equation \eqref{model-trend} and solving for \( t_{\alpha} \), we obtain:
\begin{equation}
    t_{\alpha} = \frac{-b \pm \sqrt{b^2 - 4c \left( \frac{\alpha}{100} \right) a}}{2c}
\end{equation}\label{recovery-time}

A recovery time is only meaningful if mobility initially decreased relative to the baseline, which imposes the condition \( a < 0 \). Moreover, since we are referring to a ``recovery,'' the trend must be increasing at \( t = 0 \), requiring \( y'(0) > 0 \), or equivalently, \( b > 0 \). As we are only interested in positive time values, and given \( b > 0 \), we select the solution with the \( + \) sign in equation \eqref{recovery-time}.

Finally, if \( b^2 - 4c \left( \frac{\alpha}{100} \right) a < 0 \), there is no real-valued solution for \( t_{\alpha} \). In such cases, we interpret this as mobility levels never reaching the specified recovery threshold.

The recovery time can also be obtained under a linear model (e.g. Model 5). The derivation is then analogous, but $y(t) = a + b\cdot t$ and therefore, $t_\alpha$ is given by:
\begin{equation}
    t_{\alpha} = -\dfrac{\alpha}{100}\cdot \dfrac{a}{b}
\end{equation}

\subsection{Computation of uncertainty for recovery times}

According to Model 6, the recovery time at threshold $\alpha$ is given by equation \eqref{recovery-time} where $a$, $b$ and $c$ are the coefficients for the intercept, linear and quadratic terms in equation \eqref{model-trend}.

The partial derivatives are:
\[
\frac{\partial t_\alpha}{\partial a} = \frac{-\frac{\alpha}{100}}{\sqrt{b^2 - 4c\frac{\alpha}{100} a}}, \quad
\frac{\partial t_\alpha}{\partial b} = \frac{1}{2c} \left( -1 + \frac{b}{\sqrt{b^2 - 4c\frac{\alpha}{100} a}} \right), \quad
\frac{\partial t_\alpha}{\partial c} = \frac{(b - \sqrt{b^2 - 4c\frac{\alpha}{100} a})\sqrt{b^2 - 4c\frac{\alpha}{100} a} - 2\frac{\alpha}{100} a c}{2c^2\sqrt{b^2 - 4c\frac{\alpha}{100} a}}
\]

The propagated uncertainty is:
\[
\sigma_t = \sqrt{
\left( \frac{\partial t_\alpha}{\partial a} \right)^2 \sigma_a^2 +
\left( \frac{\partial t_\alpha}{\partial b} \right)^2 \sigma_b^2 +
\left( \frac{\partial t_\alpha}{\partial c} \right)^2 \sigma_c^2}
\]

If using a linear model to estimate $t_{\alpha}$, like Model 5, then:
\[
\frac{\partial t_\alpha}{\partial a} = -\frac{\alpha}{100} \cdot \frac{1}{b}, \quad
\frac{\partial t_\alpha}{\partial b} = \dfrac{\alpha}{100}\cdot\dfrac{a}{b^2}
\]
and the propagated uncertainty is 
\begin{displaymath}
    \sigma_{t_\alpha} = \sqrt{
\left( \frac{\partial t_\alpha}{\partial a} \right)^2 \sigma_a^2 +
\left( \frac{\partial t_\alpha}{\partial b} \right)^2 \sigma_b^2 }
\end{displaymath}

\newpage

\subsection{Estimated recovery times}

\begin{table}[h!]
\centering
\includegraphics[width=0.9\linewidth]{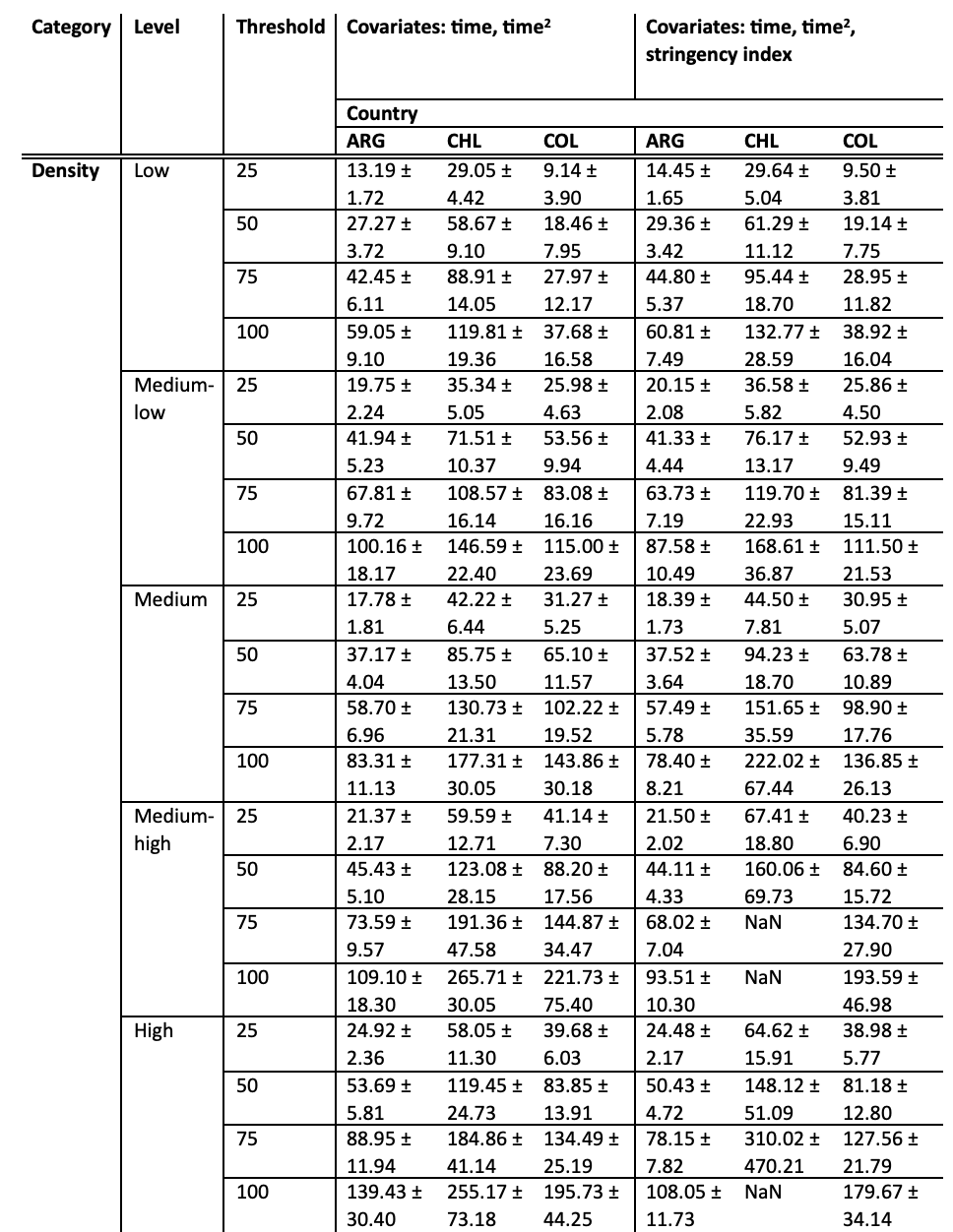}
\caption{Recovery times at multiple thresholds, for outflows in Argentina, Chile and Colombia, by population density . Computed using Model 6, with covariates time and time$^2$, or time, time$^2$ and stringency index. N/A values correspond to cases where the initial mobility levels are above the baseline. NaN values correspond to cases where mobility never recovers to the specified threshold.}
\end{table}

\begin{table}[h!]
\centering
\includegraphics[width=0.9\linewidth]{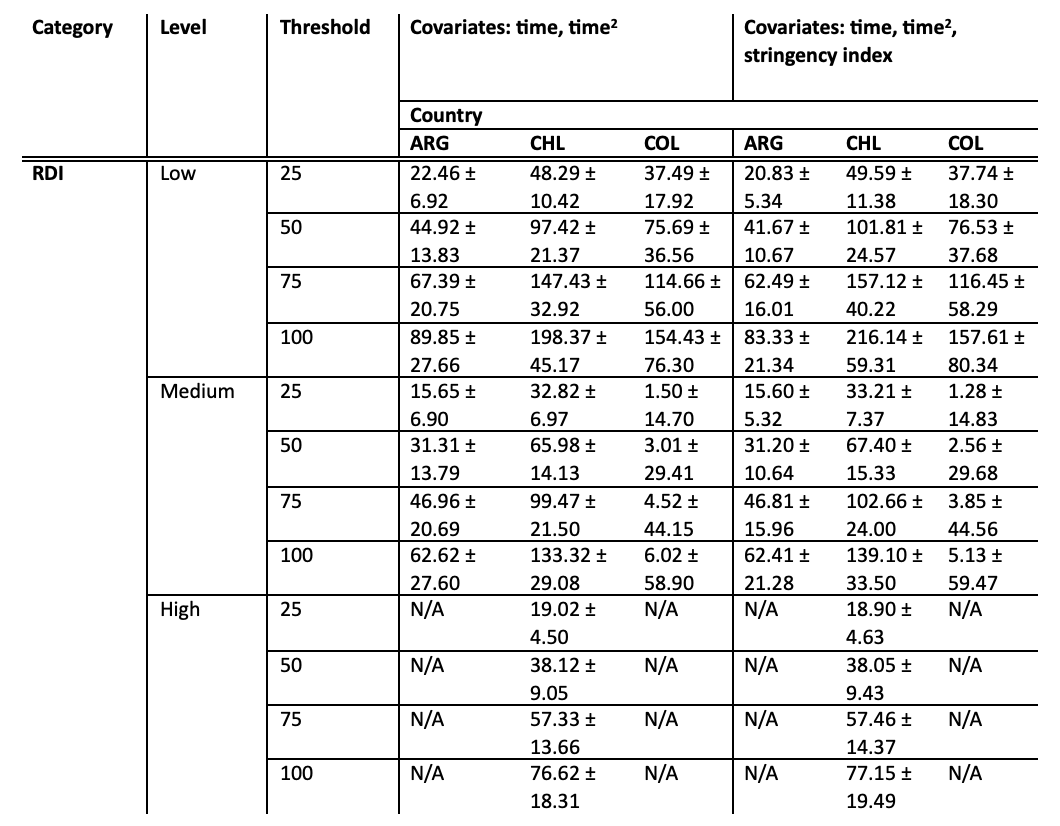}
\caption{Recovery times at multiple thresholds, for outflows in Argentina, Chile and Colombia, by Relative Deprivation Index (RDI) category. Computed using Model 6, with covariates time and time$^2$, or time, time$^2$ and stringency index. N/A values correspond to cases where the initial mobility levels are above the baseline. NaN values correspond to cases where mobility never recovers to the specified threshold.}
\end{table}

\begin{table}[h!]
\centering
\includegraphics[width=0.9\linewidth]{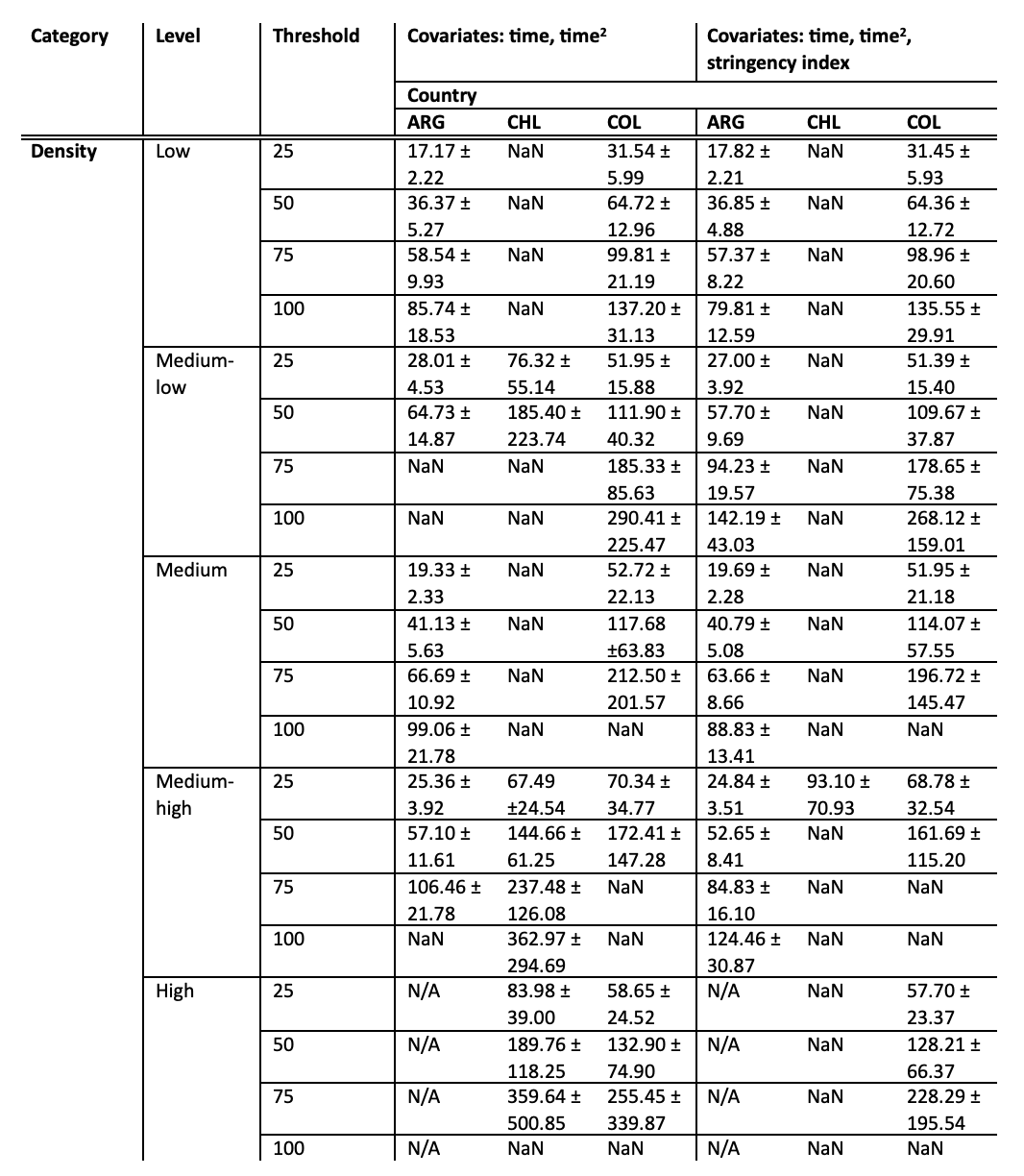}
\caption{Recovery times at multiple thresholds, for netflows involving the urban region surrounding the capital in Argentina, Chile and Colombia, by population density category. Computed using Model 6, with covariates time and time$^2$, or time, time$^2$ and stringency index. N/A values correspond to cases where the initial mobility levels are above the baseline or the only cell in a population density category is the centre of netflows. NaN values correspond to cases where mobility never recovers to the specified threshold.}
\label{tab-rec-times-capital}
\end{table}

\begin{table}[h!]
\centering
\includegraphics[width=0.9\linewidth]{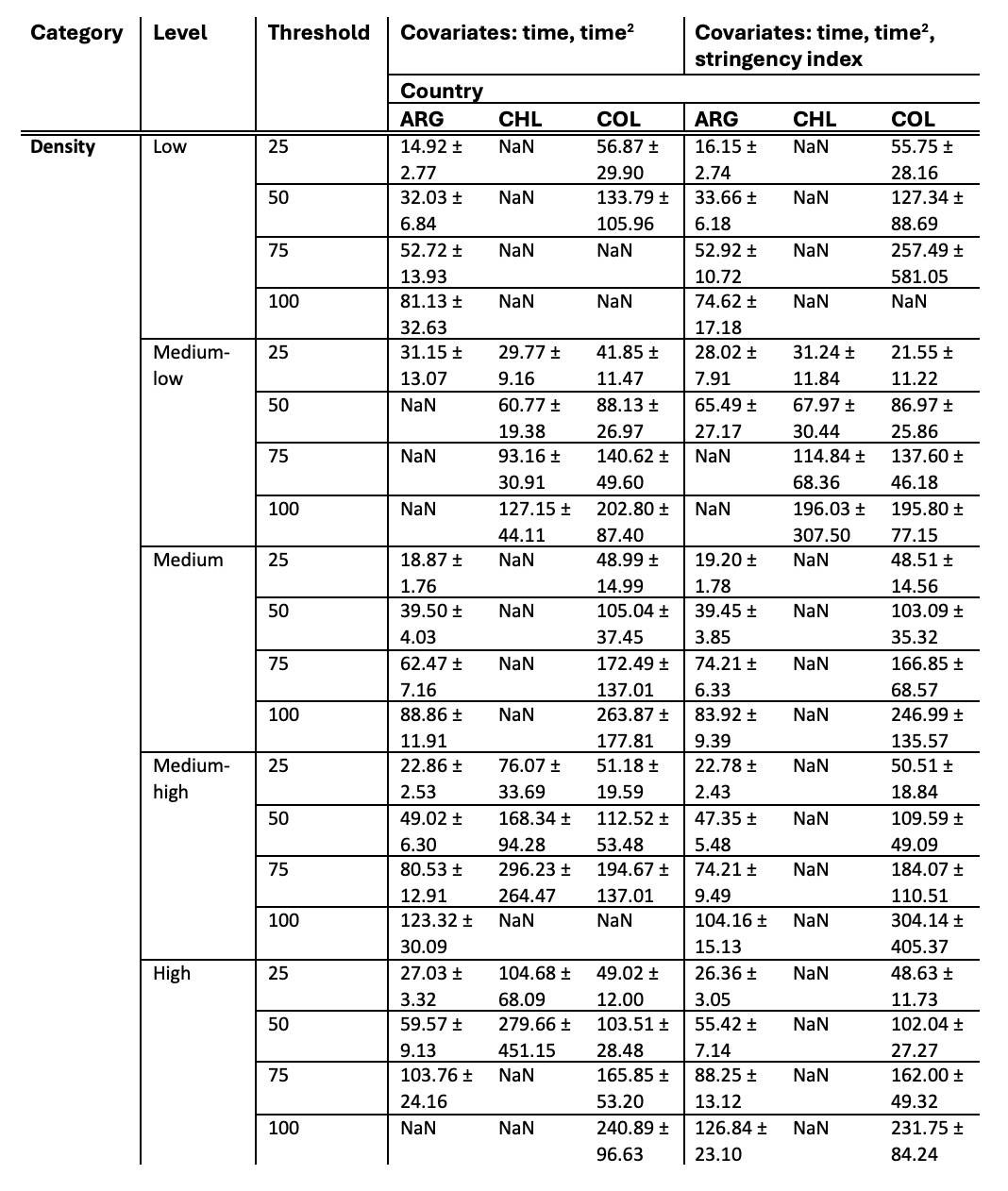}
\caption{Recovery times at multiple thresholds, for netflows involving all urban regions in Argentina, Chile and Colombia, by population density category. Computed using Model 6, with covariates time and time$^2$, or time, time$^2$ and stringency index. NaN values correspond to cases where mobility never recovers to the specified threshold.}
\label{tab-rec-times-fuas}
\end{table}